%% file: BESReview.tex
\documentclass[letterpaper]{ar-1col}
\usepackage[numbers]{natbib}
\pdfoutput=1
\usepackage[totalwidth=340pt, totalheight=570pt]{geometry}

\begin{document}

\markboth{R. A. Briere, F. A. Harris and R. E. Mitchell}{Physics Accomplishments of
  the BES Experiments}

\title{Physics Accomplishments and Future Prospects of the BES
  Experiments at the BEPC Collider}

\author{Roy A. Briere$^1$, Frederick A. Harris$^2$ and Ryan E. Mitchell$^3$
\affil{$^1$Department of Physics, Carnegie Mellon University,
  Pittsburgh, Pennsylvania, USA, 15213; email: rbriere@andrew.cmu.edu}
\affil{$^2$Department of Physics and Astronomy, The University of Hawaii,
  Honolulu, Hawaii, USA, 96734; email: fah@phys.hawaii.edu}
\affil{$^3$Department of Physics, Indiana University, Bloomington,
  Indiana, USA, 47405; email: remitche@indiana.edu}}


\begin{abstract}
The cornerstone of the Chinese experimental particle physics program 
consists of a series of experiments performed in the tau-charm energy 
region.  China began building $e^+ e^-$ colliders at the Institute 
for High Energy Physics in Beijing more than three decades ago.  
Beijing Electron Spectrometer, BES, is the common root name 
for the particle physics detectors operated at these machines.  
The development of the BES program is summarized and highlights 
of the physics results across several topical areas are presented.
\end{abstract}

\maketitle

\tableofcontents


\section{BES AND TAU-CHARM ENERGY REGION PHYSICS}

The Beijing Spectrometer experiments, BESI, BESII, and BESIII, have a
long history of operation at the Beijing Electron Positron Colliders,
BEPC and BEPCII, located at the Institute for High Energy Physics,
IHEP, in Beijing, China. BEPC and BEPCII were designed to operate in
the tau-charm center-of-mass (CM) energy region from 2 to 5 GeV.
This region provides access to a broad range of physics topics,
including charmonium and charm physics, hadron studies, determination of the tau mass, $R$ measurements, and investigations of the still-mysterious $XYZ$ particles.

\begin{figure}[h]
\includegraphics[width=3in]{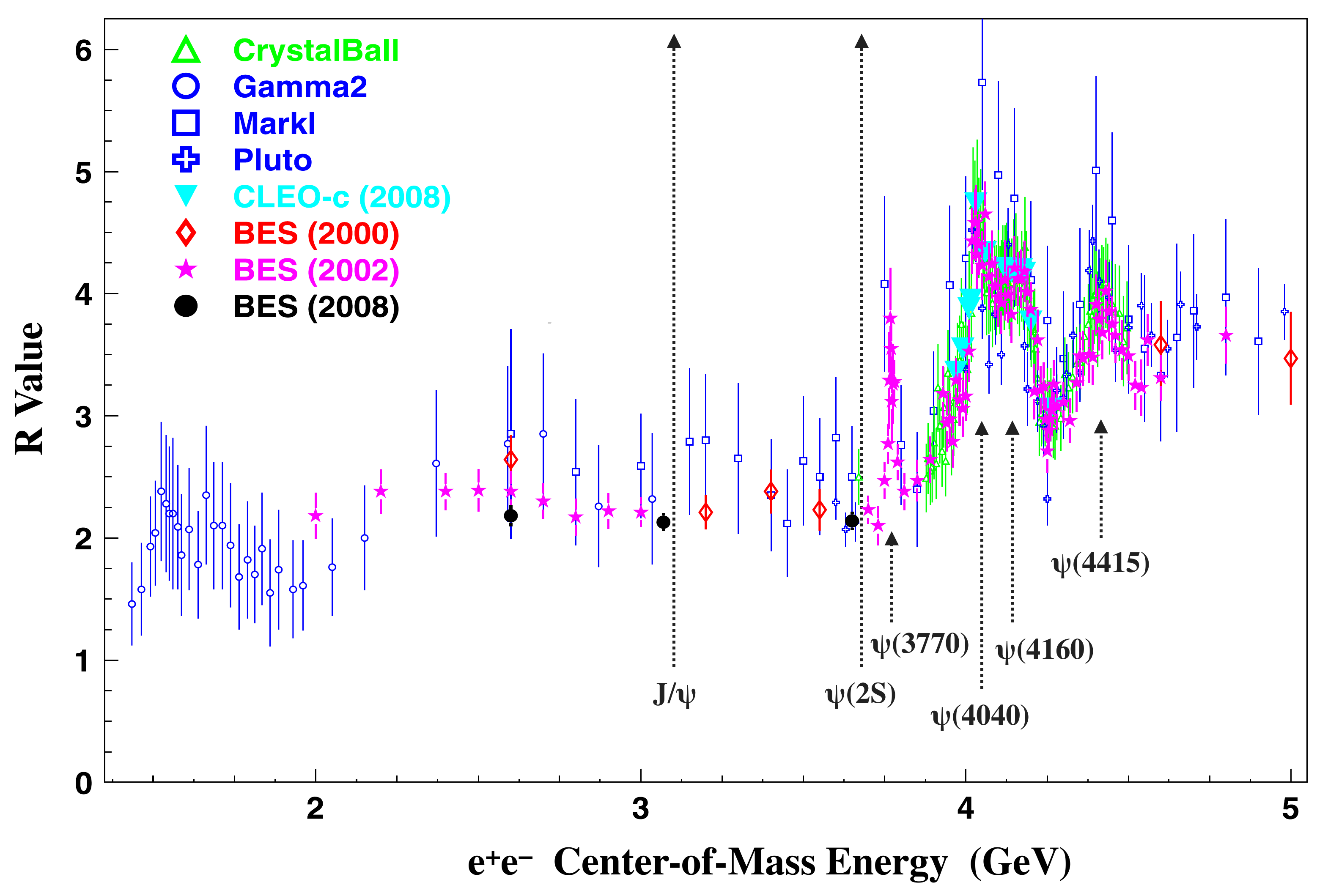}
\caption{$R = \sigma(e^+ e^- \to hadrons)/\sigma(e^+ e^- \to \mu^+
  \mu^-)$ measurements as a function of $e^+ e^-$ CM energy. Also
  shown are the positions of the $J/\psi$, $\psi(2S)$, and other
  higher mass charmonium states. Modified from
  Reference~\cite{besr2004} with permission.}
\label{Rplot}
\end{figure}

This energy region has been instrumental in understanding various
aspects of the Standard Model of elementary particle physics.
Figure~\ref{Rplot} shows the cross section for electron-positron
annihilation to hadrons divided by the cross section to muons, $R =
\sigma(e^+ e^- \to hadrons)/\sigma(e^+ e^- \to \mu^+ \mu^-)$, in the
CM energy range from 1.3 to 5.0 GeV. Except for the large $J/\psi$ and
$\psi(2S)$ charmonium resonances, the region below about 3.7 GeV is
relatively flat with an $R$ value determined approximately by the number
of kinematically accessible quark flavors (up, down, and strange) with
each quark coming in three ``colors.''  The $R$ measurements provided some
of the first evidence for ``color'' in Quantum Chromodynamics
(QCD)~\cite{riordin}.

\begin{marginnote}[4in]
  \entry{CROSS SECTION ($\sigma$)}{An effective area that, given the
    integrated luminosity, determines the likelihood an event is produced.
    Units: cm$^{2}$ or barns (1 barn = $10^{-24}$
    cm$^{2}$).}  
\end{marginnote}

The discovery of the $J/\psi$~\cite{jpsi}, composed of a charm and an
anti-charm quark ($c\bar{c}$) was instrumental in establishing the
existence of charm and in convincing physicists of the
reality of the quark model~\cite{riordin}.  The region above 4.6 GeV
is again relatively flat, but at a higher value due to crossing the
charm production threshold.

The $\psi(3770)$ is just above the threshold for producing open-charm
$D\bar{D}$ meson pairs, and it decays almost entirely to $D
\bar{D}$. A $D$ meson is formed from a charm quark and a light (up,
down) anti-quark.  The complicated region above the $\psi(3770)$ to
about 4.5 GeV is the charm meson resonance region, containing
additional $\psi$ resonances and a rich variety of other states,
including the intriguing $XYZ$ states.

Although it is not obvious from Fig.~\ref{Rplot}, the threshold for
$e^+ e^- \to \tau^+ \tau^-$ is at approximately 3.554 GeV.  The
$\tau$ lepton is the third charged lepton, in addition to the
electron and the muon.  

\begin{marginnote}[0.5in]
  \entry{LUMINOSITY($\mathcal{L}$)}{Measures
    the ``strength'' of colliding beams. Units: cm$^{-2}$ s$^{-1}$.}
  \entry{INTEGRATED LUMINOSITY ($\int{\mathcal{L}dt}$)}{The luminosity
    times the total time of the collisions.  Units: cm$^{-2}$ or
    barns$^{-1}$. Number of events expected is given by $\int{\mathcal{L}dt} \times \sigma$.}
\end{marginnote}

Remarkably, all this physics is accessible at IHEP.  BES has data sets
at many CM energies in this region and very large data sets at the
$J/\psi$ (1.3 billion events), $\psi(2S)$ (0.45 billion), and
$\psi(3770)$ (2.9 fb$^{-1}$). These are the world's largest exclusive
charmonium data sets and allow for many precision measurements.

The BES experiments have published 267 physics papers up through
the end of 2015 and have provided an innumerable number of talks and
technical papers.  In deciding what physics topics to cover here, we
have chosen to give some priority to those with the most
citations. However, we also keep in mind that many citations are made
directly to the Particle Data Group (PDG)~\cite{pdg} listings, and
that more recent papers have had less time to be cited.


\section{BEGINNINGS}

The history of the development of high-energy physics in China is
fascinating and is detailed in ``Panofsky on Physics, Politics and
Peace''~\cite{panofsky} by Wolfgang K. H. Panofsky.  In 1973,
China had decided to build a 50 GeV proton accelerator near the Ming
Tombs outside Beijing.  Panofsky was critical of this proposal since the
machine would be expensive and have less energy than similar machines
in the US and Europe.  He advised ``that an electron-positron collider
would be a much better initial venture for China, because such a
machine could serve a dual purpose of serving the economy by being a
facility for synchrotron radiation, while at the same time allowing
them to enter a field that was just beginning to be explored in the
West.''

Following much consultation, the Chinese government agreed to sponsor
the construction of the Beijing Electron-Positron Collider at IHEP.
This involved collaboration with the Stanford Linear Accelerator
Center (SLAC).  The Chinese sent a delegation of about 30 engineers
and physicists to SLAC in 1982 to make the preliminary design of the
machine.  Subsequently, the Chinese authorized construction of the
BEPC, and cooperation with SLAC continued.  Deng Xiaoping personally
wielded a shovel at the groundbreaking ceremony on Oct. 7, 1984 and
returned to IHEP on Oct. 24, 1988 to celebrate the completion of BEPC.
Important dates are summarized in Table~\ref{tab1}.

\begin{textbox}[h]
\section{Joint Committee of Cooperation in
High-Energy Physics}
In 1979, President Jimmy Carter and Chairman Deng Xiaoping signed the
United States-China Agreement on Cooperation in Science and
Technology.  The first protocol under this agreement was in
high-energy physics, and a Joint Committee of Cooperation in
High-Energy Physics (JCCHEP) has met annually since.  In 2004, it
celebrated its 25th anniversary. Panofsky and T. D. Lee, who had both
participated since 1979 and made valuable contributions, attended.
\end{textbox}

\begin{table}[h]
\tabcolsep7.5pt
\caption{Timeline}
\label{tab1}
\begin{center}
\begin{tabular}{@{}c|c|l@{}}
\hline
Dates      & Exp. & Item  \\ \hline
1979       & & First meeting of JCCHEP \\
1981       & & T.D. Lee and Panofsky suggest $e^+ e^-$ collider \\   
1982       & & Deng Xiaoping endorses $e^+ e^-$ collider \\    
4/24/1984  & & BEPC project officially approved \\             
10/7/1984  & & Ground breaking (Deng Xiaoping wields shovel)\\   
10/16/1988 & & First collisions in BEPC \\                     
10/24/1988 & & Inaugural celebration; Deng Xiaoping attends \\ 
May 1989   & BESI & BESI detector moves to interaction region \\        
6/22/1989  & BESI & $J/\psi$ peak observed in BESI \\          
Jan. 1990  & BESI & Data taking at $J/\psi$ begins \\          
May 1991   & BESI & 10 M $J/\psi$ events accumulated \\        
1991       & BESI & American scientists join; BESI becomes
international \\                                               
Nov. 1991 - Jan. 1992  & BESI & $\tau$ threshold scan \\       
1992       & BESI &  Improved $\tau$ mass measurement announced\\  
Jan. 1992 - May 1993 & BESI & $D_s$ runs (10 pb$^{-1}$) \\     
1993 - 1995 & BESI & 4 M $\psi(2S)$ accumulated \\           
1998 - 1999 & BESII & $R$-scan from 2 - 5 GeV\\  
Nov. 1999 - May 2001 & BESII & 51 M $J/\psi$ accumulated \\    
Nov. 2001 - Mar. 2002 & BESII & 14 M $\psi(2S)$ accumulated\\ 
2/14/2003 & & BEPCII approved \\                               
4/30/2004 & & BEPC shuts down and upgrade begins \\            
6/5/2005 & & First BESIII Collaboration Meeting \\             
4/30/2008 & BESIII & BESIII moves to interaction region \\     
7/18/2008 & BESIII & First hadron events recorded \\       
4/14/2009 & BESIII & 106 M $\psi(2S)$ events accumulated \\ 
7/28/2009 & BESIII & 225 M $J/\psi$ events accumulated  \\
6/27/2010 & BESIII & 0.975 fb$^{-1}$ accumulated at $\psi(3770)$   \\
5/3/2011 & BESIII & 2.9 fb$^{-1}$ accumulated at $\psi(3770)$   \\
3/31/2012 & BESIII & 0.45 B $\psi(2S)$ events accumulated \\ 
5/26/2012 & BESIII & 1.3 B $J/\psi$ events accumulated  \\
Dec. 2012 - June 2013    & BESIII & initial $XYZ$ running \\
Feb. 2014 - May 2014     & BESIII & subsequent $XYZ$ running \\
Dec. 2014 - Apr. 2015    & BESIII & $R$-scan from 2 - 3 GeV \\
\hline
\end{tabular}
\end{center}
\end{table}


\section{THE BES EXPERIMENTS}

The Beijing Electron-Positron Collider, BEPC~\cite{bepc}, originally
operated from 1988 until 1995; it was then upgraded, increasing the
reliability of the machine and approximately doubling its luminosity. %
The upgraded BEPC ran from 1998-2004, when a major upgrade to BEPCII
was started.  BEPCII is a two ring collider with 93 bunches and currents
of up to 0.91 A in each ring, and a design luminosity of $1 \times
10^{33}$ cm$^{-2}$ s$^{-1}$~\cite{bepcii}.  Some parameters of the
colliders are given in Table~\ref{BEPC}.

\begin{table}[h]
\tabcolsep7.5pt
\caption{Some BEPC and BEPCII parameters}
\label{BEPC}
\begin{center}
\begin{tabular}{@{}l|c|c|c@{}}
\hline
Parameter                      & BEPC & Upgrade & BEPCII         \\
\hline
Beam energy  (GeV)             &  1.1 - 2.7 & 1.0 - 2.8  & 1.0 - 2.3     \\
Design luminosity ($ \times 10^{33}$ ) (cm$^{-2}$ s$^{-1}$) & $0.0065$ & N/A   & $1 $ \\
~~~~~~~~at beam energy (GeV)           & 2.2     &  N/A    & 1.89            \\
Obtained luminosity ($\times 10^{33}$) (cm$^{-2}$ s$^{-1}$) & $0.007$  & $0049
$& $0.853$ \\
~~~~~~~~at beam energy (GeV)           &  2.2    & 1.55   & 1.89            \\
No. bunches                    &  1   &  1   & 93              \\
Beam current (A)              & 0.03  &  0.045 & 0.91 (Nominal)  \\
~~~~~~~~at beam energy (GeV)  & 2.2   &  1.55  &             \\
Circumference (m)              & 240 & 240 & 237             \\
\hline
\end{tabular}
\end{center}
\end{table}

The configurations of the BES detectors are similar, although
the subsystems themselves are often quite different. For all three,
the innermost subsystem is composed of drift chamber(s) to determine
the momenta and trajectories of charged particles in the magnetic
field. Next are time of flight (TOF) counters to determine their
velocities, followed by electromagnetic shower counters to measure the
energies of photons and identify electrons. Outside the
electromagnetic shower counter is the coil of the magnet with the flux
return instrumented with detectors to identify muons by their
penetration through the iron.

BESI had a central drift chamber (CDC) surrounded by the main drift
chamber (MDC).  Its electromagnetic calorimeter was composed of self
quenching streamer tubes interleaved with lead.  Details of BESI may
be found in Ref.~\cite{BESI}.  BESI operated from 1989 until 1995,
when it was upgraded to BESII, and BEPC was also upgraded.  The upgrade
replaced the CDC with a revamped MARKII vertex detector and replaced
the MDC and the barrel TOF system.  BESII operated from 1998 until
2004. Details on BESII may be found in Ref.~\cite{BESII}.

\begin{figure}[h]
\includegraphics[height=2.0in]{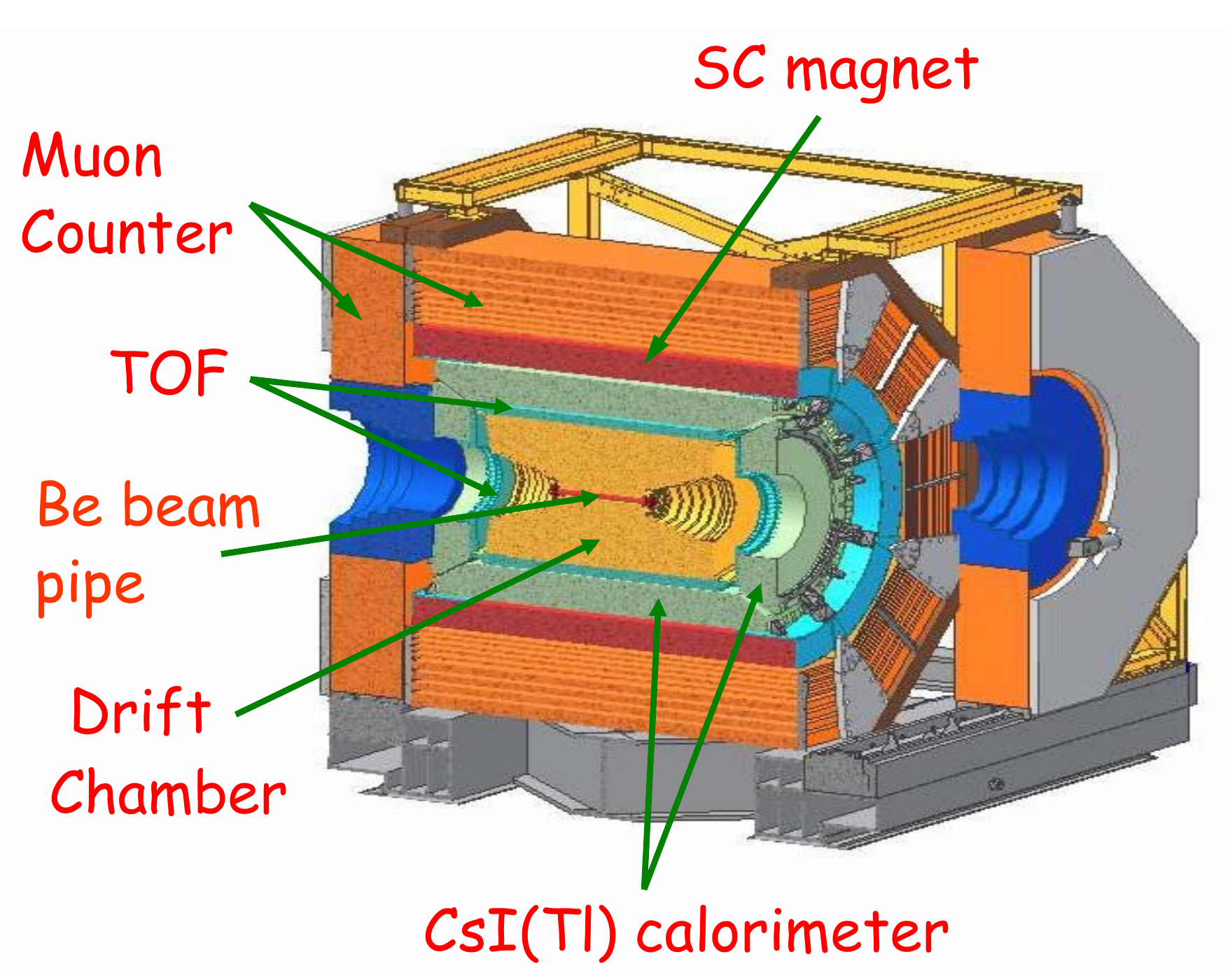}
\caption{Schematic of BESIII detector.  Shown are the beryllium beam
  pipe, main drift chamber, barrel and end-cap TOF counters, the barrel
  and end-cap CsI(Tl) electromagnetic calorimeters, the 1 T
  superconducting magnet, and the muon resistive plate chambers
  embedded in the magnet return yoke iron. The outer radius of the
  main drift chamber is 0.81 m.}
\label{bes3detector}
\end{figure}

The current detector is BESIII, which is a new detector with a single
small-celled, helium-based MDC, a plastic scintillator TOF system, a
CsI(Tl) electromagnetic calorimeter, a 1.0 T superconducting magnet,
and a muon counter with 9 resistive plate chamber (RPC) layers in the
barrel part and 8 in the end-cap portions interleaved in the steel of
the flux return yoke.  Details of BESIII may be found in
Ref.~\cite{BESIII}. Figure~\ref{bes3detector} shows a schematic view
of the BESIII detector, and some details on all three detectors are
summarized in Table~\ref{detectors}.

\vspace{1in}

\begin{table}[h]
\tabcolsep7.5pt
\caption{Some BES detector parameters.  }
\label{detectors}
\begin{center}
\begin{tabular}{@{}l|c|c|c|c@{}}
\hline
Sub-system &  Parameter        & BESI & BESII& BESIII                 \\
\hline
Beam pipe  &   Material               &  Al  &      & Be    \\ \hline
MDC        & \# layers         & 40    & 40  & 43   \\
           & $\sigma_p/p$ @ 1~GeV/$c$ & 2.4\% & 2.5\% & 0.5\%    \\
           & $\sigma_{dE/dx}$ & 8.5\%    & 8\%  & 6\% @ 1~GeV/$c$  \\ \hline
TOF-barrel & \# scint.         & 48     & 48 & 2 layers/88 in each  \\
           & $\sigma_t$       & 330 ps    &180 ps   & 80 ps                \\ \hline
TOF-end-cap     & \# scint. (each end) & 24  &  24    &  48        \\
           &  $\sigma_t$      & ND     &  ND    &   110 ps     \\ \hline
EMC-barrel &  Construction    & Str. tubes/Pb &  Str. tubes/Pb  & CsI(Tl)    \\
           &  $\sigma_E/E$ @ 1~GeV  & 24\% &  21\%    & 2.5\%           \\
           &  $\sigma_{pos}$ (cm)  &  3.0    & ND   & 0.6  \\
\hline
EMC-end-cap     &  Construction   & Str. tubes/Pb & Str. tubes/Pb & CsI(Tl)    \\
           &  $\sigma_E/E$ @ 1~GeV  & 21\% &   21\%    &   5\%   \\
           &  $\sigma_{pos}$ (cm)  & 2.3     &  2.3   &  0.9 \\
\hline
Magnet     &  Type           & conventional    & conventional  & superconducting      \\
           &  Field  (T)      & 0.4  & 0.4    & 1                    \\ \hline
Muon-barrel&  \# layers        & 3     &  3  & 9  RPCs     \\ 
           &  $\sigma_{pos}$ (cm) &  6   &  6   &   2\\
 \hline
Muon-end-cap   &  \# layers        &  ND    &  ND    & 8  RPCs       \\ 
\hline
\end{tabular}
\end{center}
\end{table}


\section{\boldmath $\tau$ MASS MEASUREMENTS}

In the early 1990's, the $\tau$ lepton appeared to violate the Standard Model.
According to theory, the $\tau$ lifetime ($\tau_{\tau}$), $\tau$ mass
($m_{\tau}$), electronic branching fraction ($B(\tau\rightarrow
e\nu\bar{\nu})$) and weak coupling constant $g_{\tau}$ are related to
one another according to:
\begin{eqnarray}
 \frac{B(\tau\rightarrow e\nu\bar{\nu})}{\tau_{\tau}}=\frac{g_{\tau}^2
 m_{\tau}^5}{192\pi^3},
\label{Eq.calgtau}
\end{eqnarray}
up to small radiative and electroweak corrections~\cite{LepUnivCor}.
However, this relation appeared to be badly violated, and BES/BEPC was
in an excellent position to measure the $\tau$ lepton mass, one of the
fundamental parameters of the Standard Model.

In spring 1992, the BES collaboration, composed then of more than 100
Chinese physicists from IHEP and about 40 American physicists, measured
the mass to be $1776.9 ^{+0.4}_{-0.5} \pm 0.2$~MeV/$c^2$ by an energy scan
over the $\tau$ production threshold using the reaction $e^+ + e^- \to
\tau^+\tau^- \to e^{+} \nu_{e}\bar{\nu_{\tau}} \mu^-
\bar{\nu_{\mu}}\nu_{\tau}$~\cite{bes1}.  Approximately 5
pb$^{-1}$ of data, distributed over 12 scan points, was collected.  The 
mass was lower than the world average value at that time by 7.2~MeV/$c^2$,
had improved precision by a factor of 7, and greatly improved
agreement with the Standard Model.
This measurement was later updated to be $1776.96
^{+0.18+0.25}_{-0.21-0.17}$~MeV/$c^2$ with more $\tau$
decay channels~\cite{bes2prd}.


The new BESIII detector and BEPCII accelerator called for an improved
$\tau$ mass measurement. A study was carried out before starting a new
energy scan to optimize the number and choice of scan points in order
to provide the highest precision for a given integrated
luminosity~\cite{optstd3}.

\begin{textbox}[h]
\section{Beam Energy Measurement System}
Extremely important in the threshold scan is to precisely determine the beam
energy and the beam energy spread.  For this, the beam energy
measurement system (BEMS)~\cite{bems} for BEPCII was used.  Photons
from a CO$_2$ laser are collided head on with either the electron or
the positron beam, and the maximum energies of the back scattered
Compton photons are measured with high accuracy by a High Purity
Germanium (HPGe) detector, whose energy scale is calibrated with
photons from radioactive sources.  The beam energies can be determined
by the kinematics of Compton scattering~\cite{Rullhusen}.  
\end{textbox}

The $\tau$ scan experiment was done in December 2011.  For energy
calibration purposes, the $J/\psi$ and $\psi^{\prime}$ resonances were
each scanned at seven energy points.  About $24$ pb$^{-1}$ of data,
distributed over four scan points near $\tau$ pair production
threshold, was collected.  The first point was below the mass of
$\tau$ pairs, while the other three were above.  However,
running conditions were not optimal, so the running was stopped before
collecting the full data set.

\begin{figure}[!htb]
\begin{center}
\begin{minipage}[t]{2.4in}
\includegraphics[height=4.1cm]{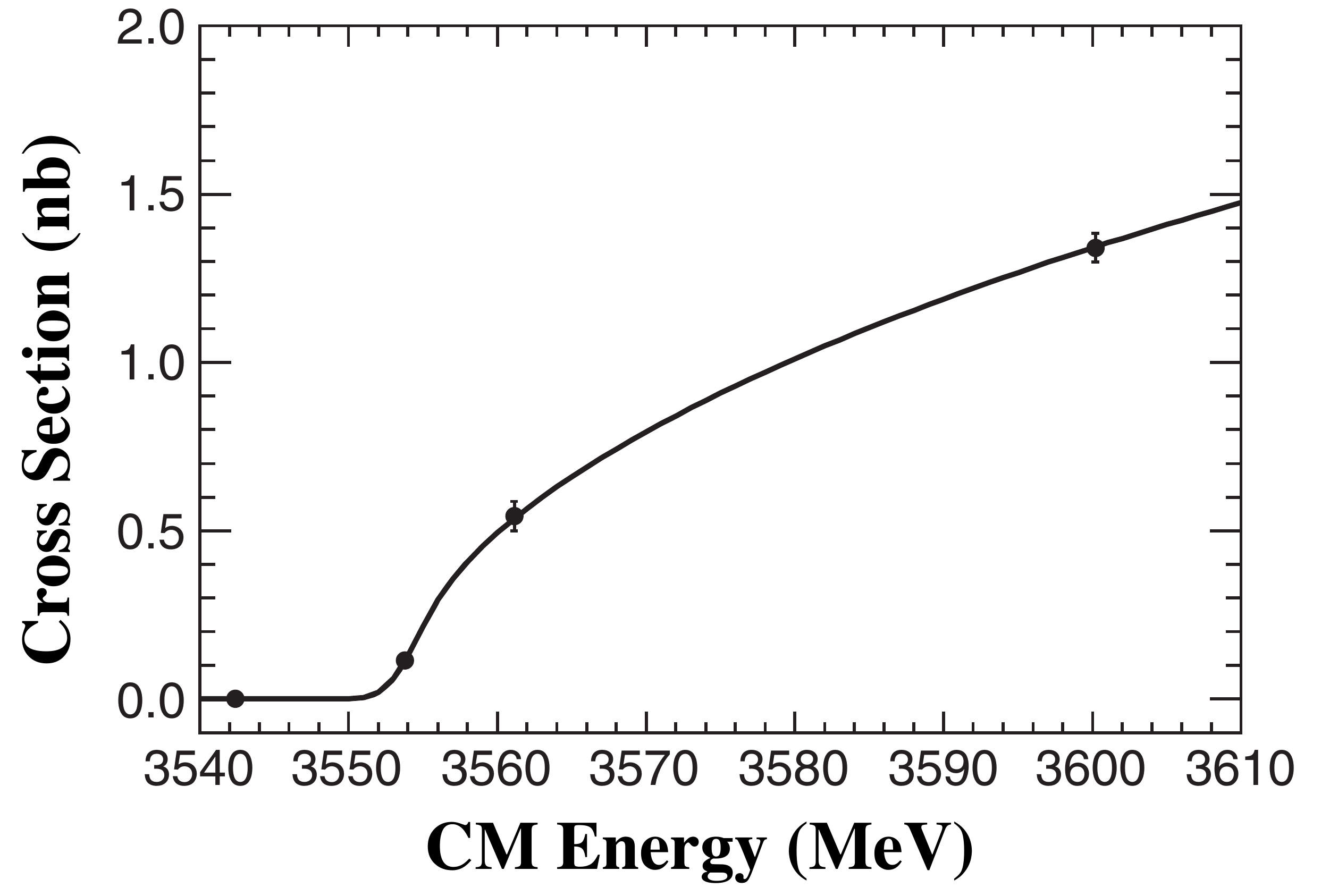}
\end{minipage} \ \
\begin{minipage}[t]{2.40in}
\includegraphics[height=4.1cm]{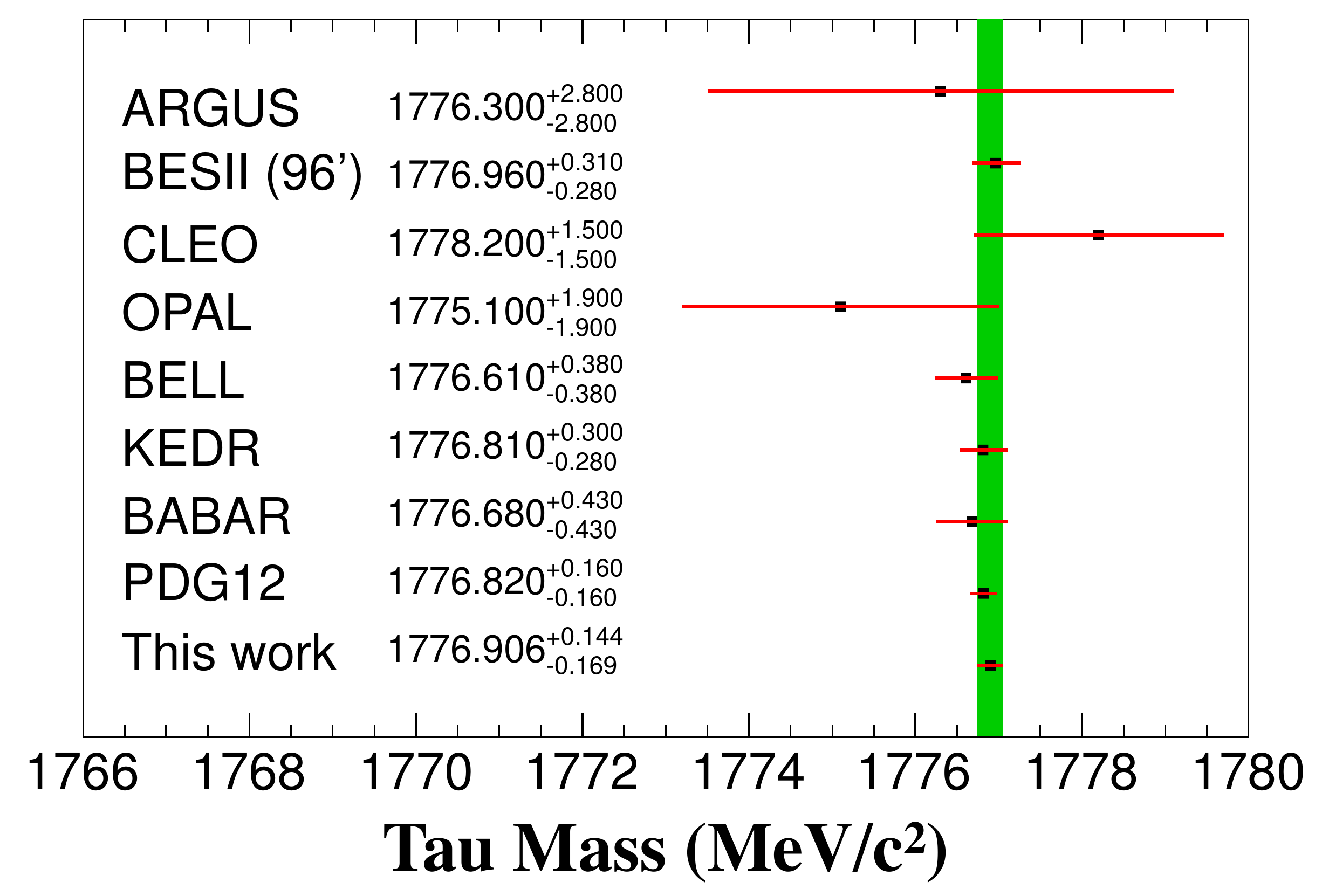}
\end{minipage}
\caption{\label{taumass_figures} (Left plot) Cross section versus $e^+
  e^-$ CM energy. Cross section measurements are shown with error
  bars; the smooth curve is the fit.  (Right plot) Comparison of the
  measured $\tau$ mass with those from the PDG~\cite{pdg}. The green band
  corresponds to the 1~$\sigma$ limit of the BESIII measurement.  Modified from
Reference~\cite{taubes3} with permission.}
\end{center}
\end{figure}

To reduce the statistical error in the $\tau$ lepton mass, the
analysis included 13 $\tau$ pair final states decaying into two
charged particles ($ee$, $e\mu$, $e\pi$, $eK$, $\mu\mu$, $\mu\pi$,
$\mu K$, $\pi K$, $\pi\pi$, $KK$, $e\rho$, $\mu\rho$ and $\pi\rho$)
plus accompanying neutrinos to satisfy lepton conservation.  By a fit
to the $\tau$ pair cross section data near threshold, shown in the left
plot of Fig.~\ref{taumass_figures}, the mass of the $\tau$ lepton was
determined to be~\cite{taubes3}:
\begin{eqnarray}
\label{finaltaumass} m_{\tau} = (1776.91\pm0.12^{+0.10}_{-0.13})~\mathrm{ MeV}/c^2 .
\end{eqnarray}
The right plot of Fig.~\ref{taumass_figures} shows the comparison of
this result with values from the PDG;
it is consistent with all of them, but has the smallest uncertainty.
With the full $\tau$ scan data set, BESIII should be able to do even
better.

\begin{textbox}[h]
\section{Lepton Universality}
A precision $m_{\tau}$ measurement is also required to check lepton
universality. Lepton universality, a basic ingredient in the minimal
Standard Model, requires that the charged-current gauge coupling strengths for the
electron, muon, and tau leptons, $g_{e}$, $g_{\mu}$, $g_{\tau}$,
should be identical: $g_e=g_{\mu}=g_{\tau}$. Lepton universality
implies:
\begin{eqnarray}
    \left(\frac{g_{\tau}}{g_{\mu}}\right)^{2}=\frac{\tau_{\mu}}{\tau_{\tau}}\left(\frac{m_{\mu}}{m_{\tau}}\right)^{5}\frac{B(\tau\rightarrow
            e\nu\bar{\nu})}{B(\mu\rightarrow
                e\nu\bar{\nu})}(1+F_{W})(1+F_{\gamma}) = 1,
\label{Eq.universaltest}
\end{eqnarray}
where $F_{W}$ and $F_{\gamma}$ are the weak and
electromagnetic radiative corrections~\cite{LepUnivCor}.  Note
$(g_{\tau}/g_{\mu})^2$ depends on $m_{\tau}$ to the fifth power.

Inserting the $\tau$ mass value into Eq.~\ref{Eq.universaltest},
together with the values of $\tau_{\mu}$, $\tau_{\tau}$, $m_{\mu}$,
$m_{\tau}$, $B(\tau\rightarrow e\nu\bar{\nu})$ and $B(\mu\rightarrow
e\nu\bar{\nu})$ from the PDG~\cite{pdg} and using the values of
$F_{W}$ (-0.0003) and $F_{\gamma}$ (0.0001)) calculated from
reference~\cite{LepUnivCor}, the ratio of squared coupling constants
is determined to be:
\begin{eqnarray}
\label{Eq.lepuniversal} \left(\frac{g_{\tau}}{g_{\mu}}\right)^{2}=1.0016\pm0.0042,
\end{eqnarray}
which is consistent with unity.
\end{textbox}


\section{R SCAN}

The big news of 2012 was the discovery of the Higgs particle, the
capstone of the Standard Model, at the Large Hadron Collider (LHC) at CERN in
Geneva, Switzerland.  However, before the discovery, fits in the Standard Model
were able to predict its mass, because higher order terms in the model
can include a massive virtual particle, such as the Higgs.
Surprisingly, important in this fit is $R$ scan data.

Among the three input parameters generally used in global fits to
electroweak data, the QED running coupling constant evaluated at the
mass of the $Z$ boson, $\alpha(M^2_{Z})$, has the largest experimental
uncertainty.  While its value at low energy, $\alpha(0)$, is known
precisely, the correction necessary to determine its value at high
energy, $\alpha(M^2_{Z})$, cannot be reliably calculated
theoretically.  Instead, experimentally measured $R$ values are used
with the application of dispersion relations~\cite{rlowe}.

Uncertainties in the values of $R$ limit the precision of
$\alpha(M^2_{Z})$, which in turn limits the precision of the
determination of the Higgs mass~\cite{6,7,8}.  Before the measurement
by BESII, the uncertainty in $\alpha(M^2_{Z})$ was dominated by the
errors of the values of $R$ in the CM energy range below 5 GeV.
These were measured about 20 years earlier with a precision of about
$15\sim 20\%$ and accounted for about 50\% of the uncertainty in
$\alpha(M^2_{Z})$~\cite{9}.  With these $R$ values, the best fit value
for the Higgs' mass was $M_{H}=62^{+53}_{-30}$~GeV/$c^2$~\cite{9},
which was about one standard deviation below the lower limit of
$M_{H}>114$~GeV/$c^2$ coming from experiments at the CERN Large Electron
Positron (LEP) collider~\cite{heister}.  However, the calculated
result was very sensitive to the value used for $\alpha
(M_{Z}^{2})$. Clearly, a more precise determination $\alpha
(M_{Z}^{2})$ was very important.


In 1998 and 1999, $R$ value measurements were made at 91 energy
points~\cite{besr98,besr99} between 2 and 5~GeV by BESII. The BESII
$R$ values are displayed in Fig.~\ref{Rplot} along with those from
other experiments.  BESII systematic uncertainties are between 6 and
10 \% with an average uncertainty of 6.6 \% and are a factor of two to
three improvement in precision in the 2 to 5~GeV energy region.
Ref.~\cite{besr99} is the second most highly cited BES paper with 289
citations through the end of 2015.

The importance of these results has been emphasized by Burkhardt and
Pietryzk~\cite{burkhardt}.  With the new BES $R$-values, they obtained
a value for $\alpha ^{-1}(M_{Z}^{2}) = 128.936 \pm 0.046$, where the
error is about one-half the 1995 error.  The CERN Electroweak Group
found that this result shifted the central value for the Higgs mass
upward to $M_{H}=98$~GeV/$c^2$, which was in better agreement with the LEP
lower limits.  The measured mass from the LHC of the Higgs boson is
125~GeV/$c^2$~\cite{pdg}.  


In 2004, large-statistics data samples were accumulated by BESII at CM
energies of 2.60, 3.07 and 3.65~GeV; the total integrated luminosity
was 10.0 pb$^{-1}$~\cite{besr2004}.  Improvements in the event
selection, luminosity measurement, and the use of a
GEANT3-based~\cite{GEANT3A, GEANT3B} simulation were made in order to
decrease the systematic errors. With these improvements, the errors on
the new measured $R$ values were reduced to about $3.5\%$.  These
$R$-values are also shown in Fig.~\ref{Rplot}.


BESIII has also made $R$ scans.  In 2014, a fine scan of 104 energy
points through the resonance region above 3.8~GeV was done.  The total
data accumulated was 0.8 fb$^{-1}$, which will be used to determine
$R$, study $XYZ$ particles, study the $\Lambda_c$, etc.  In 2015, 20
points were scanned in the continuum region from 2.0~GeV to 3.1~GeV.
These data will be used to determine $R$, determine baryon form
factors, and study baryon threshold behavior.  Stay tuned for the results.


\section{LIGHT QUARK PHYSICS}

\label{sec:lightQuark}

The study of light quark mesons and baryons (mesons and baryons composed of up, down, and strange quarks) has been a major aspect of each incarnation of the BES experiment.  
Charmonium states, such as the $J/\psi$, decay to hundreds of different combinations of light quark hadrons, like $\pi^+\pi^-\pi^0$, $K^+K^-\pi^+\pi^-$, and $\gamma \pi^0 \pi^0$, to name just a few.
This provides many opportunities to identify intermediate ``resonances'' in the decay sequences.  For example, in the decay $J/\psi \to \gamma \pi^0 \pi^0$, one can search for the intermediate process $J/\psi \to \gamma f_0(1710)$, with the $f_0(1710)$ subsequently decaying to $\pi^0\pi^0$, and thereby learn about the $f_0(1710)$ isoscalar state~(discussed below).  Furthermore, since the quantum numbers of the initial charmonium state are known, conservation rules can be used to derive amplitudes describing the behavior of the decay products under different assumptions about their quantum numbers.  These quantum mechanical amplitudes can be added coherently and then squared, leading to distributions that can be fit to data.  Comparing fits, and comparing the strengths of different amplitudes within the fits, one can then distinguish between different hypotheses about the quantum numbers of the final state.  This process, referred to as partial wave analysis~(PWA),  is an important aspect of the light quark physics program at BES.

\begin{marginnote}[0.5in]
  \entry{ISOSPIN}{A quantum number describing the configuration
    of up and down quarks within a hadron.}
  \entry{ISOSCALAR}{A hadron with zero units of isospin.}
\end{marginnote}

Since the $e^+e^- \to J/\psi$ cross section is so large, and since the $J/\psi$ decays predominantly to light quark states, the $J/\psi$ is the charmonium state most often used by BES to study light quark mesons and baryons.  Thus, within the collaboration, ``light quark physics" is almost synonymous with ``$J/\psi$ physics."
From BESI to BESIII, the size of the $J/\psi$ data set has grown over two orders of magnitude.  BESI collected a sample of 8.6~million $J/\psi$ decays; BESII collected 58~million; and BESIII took an initial sample of 225~million~(in 2009), and subsequently increased it to 1.3~billion $J/\psi$ decays~(in 2012).

The following sections include a few high-profile examples of how light quark mesons and baryons have been studied in $J/\psi$ decays at BES.  But it should be noted that there are other interesting physics topics not discussed here, such as the physics of $\eta$ and $\eta^\prime$ decays (which can be produced cleanly in $J/\psi$ decays).

\begin{textbox}[h]
\section{Mesons and Baryons}
Hadrons, or particles that interact via the strong force, are broadly classified by their total spin.  Mesons have integral spin; baryons have fractional spin.  The majority of mesons that have been discovered can be neatly described using a model in which they are composed of a quark and an antiquark.  Similarly, most baryons can be successfully described as composites of three quarks.  The exceptions are particularly interesting since they could represent novel configurations of matter, such as four-quark mesons (tetraquarks), or five-quark baryons (pentaquarks).  Configurations such as these are allowed in QCD, but their properties are a subject of intense experimental investigation.
\end{textbox}

\subsection{Glueballs and the light isoscalar spectrum}


One of the most high-profile aspects of light quark spectroscopy at BES is the search for glueballs in radiative $J/\psi$ decays.  Glueballs are states composed of gluons (containing no valence quarks) and their existence is a prominent prediction of QCD~\cite{Morningstar:1999rf}.  Their identification requires comparing their rate of production in different environments~\cite{Crede:2008vw}.  They should not be heavily produced in $\gamma\gamma$ collisions, for example, since there is no coupling between photons and gluons. On the other hand, the production of glueballs is expected to be enhanced in radiative $J/\psi$ decays.  In this process, the charm or anti-charm quark of the $J/\psi$ first radiates a photon, leaving the charm and anti-charm quark pair to subsequently annihilate into two gluons, which then hadronize.  Such a ``glue-rich'' environment is expected to be favorable for glueball production.


A lot of attention was garnered at BESI~\cite{Bai:1996wm} (and other contemporaneous experiments) for the apparent confirmation of a spin-2 glueball candidate, the $\xi(2230)$, first reported by MARKIII~\cite{Baltrusaitis:1985pu}.  It was seen to appear in many $J/\psi$ radiative decays, including 
$\gamma\pi^+\pi^-$~(4.6$\sigma$ evidence),
$\gamma K^+K^-$~(4.1$\sigma$ evidence),
$\gamma K_S^0 K_S^0$~(4.0$\sigma$ evidence), and
$\gamma p\bar{p}$~(3.8$\sigma$ evidence).
It had several properties that made it an ideal glueball candidate:  its mass was 
consistent with the mass expected for the tensor glueball;  it decayed in a 
``flavor-symmetric'' pattern;  it was anomalously narrow.  Unfortunately, this state was not subsequently confirmed by the BESII or BESIII collaborations, and it appears to have been an extremely unlucky fluctuation.
Since that time, there has been no observed state whose properties have made it such an appealing candidate for a glueball state.


The most promising place to look for glueballs is currently in the isoscalar spectrum, where there is an overpopulation of reported states.  If all mesons were composed of a quark and anti-quark, there would be two isoscalar states, one a mixture of up and down quarks (the $n\bar{n}$ state) and one composed of strange quarks (the $s\bar{s}$ state).  Instead, three states are seen, namely the $f_0(1370)$, $f_0(1500)$, and $f_0(1710)$.  This could indicate that one of these states is a glueball.  Unfortunately, mixing is also allowed among these states, complicating this picture~\cite{Close:2000yk}.  Thus, the $f_0(1500)$, say, could be partly $n\bar{n}$ and partly glueball, and so on.  BES has added a tremendous amount of information related to this problem, a sampling of which is included below.  Even so, a final solution has yet to be found and work continues.

\begin{figure*}
\begin{center}
\includegraphics[width=1.0\linewidth]{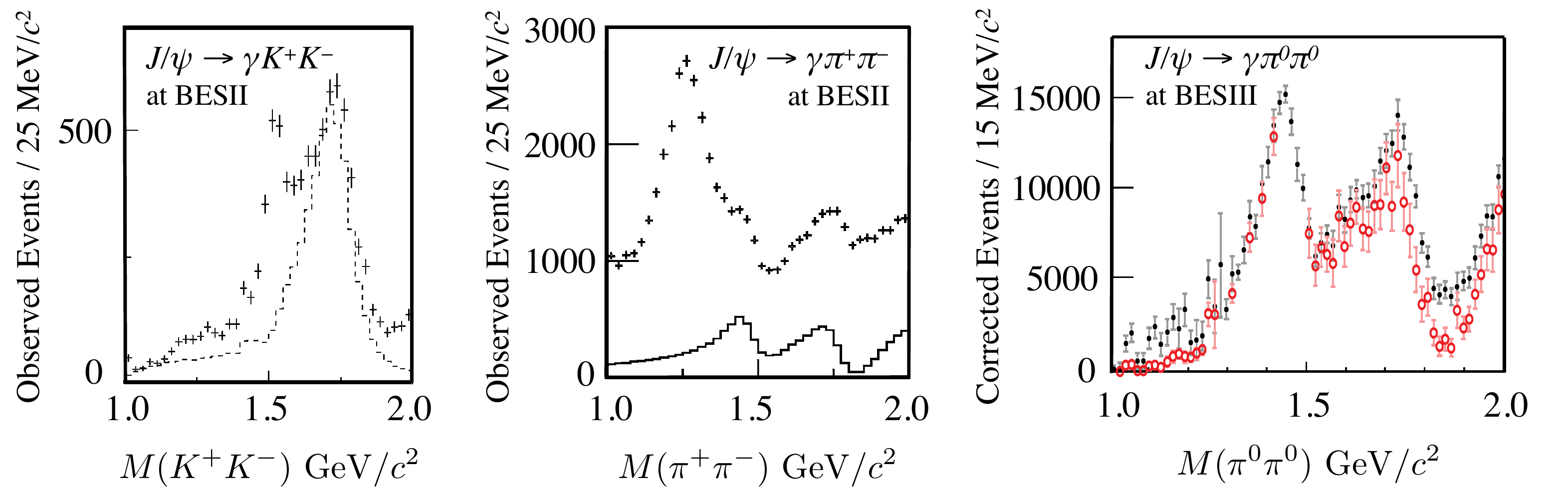}
\caption{A few representative analyses of $J/\psi$ radiative decays at BES. (left)~Analysis of $J/\psi \to \gamma K^+ K^-$ at BESII~\cite{Bai:2003ww}.   The points are data and the histogram shows the spin-0 components of the fit to data.  The peak around 1.7~GeV/$c^2$ is from the $f_0(1710)$. (middle)~Analysis of $J/\psi \to \gamma \pi^+ \pi^-$ at BESII~\cite{Ablikim:2006db}.  The points are data and the histogram shows the spin-0 components of the fit to data.  The peak just under 1.5~GeV/$c^2$ is due to the $f_0(1500)$, while the peak around 1.7~GeV/$c^2$ is from the $f_0(1710)$. (right)~Analysis of $J/\psi \to \gamma \pi^0 \pi^0$ at BESIII~\cite{Ablikim:2015umt}.  The points show the spin-0 components of the fits done in each mass bin. The fits make no assumption about the mass-dependence of the amplitudes, but new complications are thereby introduced. The solid~(black) and hollow~(red) points are mathematically ambiguous solutions.  Modified from References~\cite{Bai:2003ww,Ablikim:2006db,Ablikim:2015umt} with permission.}
\label{fig:JpsiRad}
\end{center}
\end{figure*}

The first major contribution of BES to the isoscalar problem was in the clarification of the spin of the $f_0(1710)$.  This state, discovered by the Crystal Ball experiment in 1982~\cite{Edwards:1981ex}, was initially thought to be spin-2.  BESI observed the $f_0(1710)$ produced prominently in the reaction $J/\psi \to \gamma K^+ K^-$~\cite{Bai:1996dc}.  Analyzing its decay to $K^+K^-$, BESI reported that, instead of being purely spin-2, it was actually a mixture of spin-0 and spin-2.
Later, with the increase of $J/\psi$ decays between BESI and BESII, BESII was able to do a reanalysis of the $J/\psi \to \gamma K^+ K^-$ reaction, also adding the related $J/\psi \to \gamma K_S^0 K_S^0$ process~\cite{Bai:2003ww}.  With the much increased statistics, the $f_0(1710)$ was conclusively identified as spin-0, agreeing with other contemporaneous experiments, and now the accepted value.  The results of this analysis are shown in the left panel of Figure~\ref{fig:JpsiRad}.  The $J/\psi \to \gamma K\bar{K}$ channel is still being analyzed at BESIII.


Another major contribution from BES was the analysis of $J/\psi \to \gamma \pi \pi$.  This was first performed at BESI with low statistics~\cite{Bai:1998tx}, but later studied more conclusively at BESII, using both $\pi^0\pi^0$ and $\pi^+\pi^-$~\cite{Ablikim:2006db}.  Here, both the $f_0(1500)$ and $f_0(1710)$ were seen (see the middle panel of Figure~\ref{fig:JpsiRad}).  This allowed a number of conclusions to be drawn.  First, since the $f_0(1500)$ was not seen in $J/\psi \to \gamma K\bar{K}$ its decay to $K\bar{K}$ must be significantly smaller than its decay to $\pi\pi$.  Second, the rate of production of the $f_0(1710)$ could also be compared to the previous $K\bar{K}$ analyses.  From this, the ratio $B(f_0(1710)\to \pi\pi)/B(f_0(1710)\to K\bar{K})$ could be derived to be $0.41^{+0.11}_{-0.17}$.  This remains the best measurement of this branching ratio.  Since the $f_0(1500)$ decays more often to $\pi\pi$ than $K\bar{K}$ it is more likely to be the $n\bar{n}$ state than the $s\bar{s}$ state.  And conversely, the $f_0(1710)$ is more likely to be the $s\bar{s}$ state.


In principle, these isoscalar states could also be studied by looking at how they are produced alongside the $\omega$ and $\phi$ in the four reactions $J/\psi\to\omega K^+K^-$, $\omega \pi^+\pi^-$, $\phi K^+K^-$, and $\phi \pi^+\pi^-$.  Since the $\phi$ is an $s\bar{s}$ state, it is expected, for example, that the $f_0(1710)$ is more likely to be produced alongside it than the $\omega$, which is an $n\bar{n}$ isovector.  All four reactions were studied at BESII~\cite{Ablikim:2004wn,Ablikim:2004qna,Ablikim:2005ni}, but surprisingly, the opposite was found.  While $B(J/\psi \to \phi f_0(1710))\times B(f_0(1710)\to K^+K^-)$ was measured to be $(2.0 \pm 0.7) \times 10^{-4}$, $B(J/\psi \to \omega f_0(1710))\times B(f_0(1710)\to K^+K^-)$ was measured to be $(6.6 \pm 1.3) \times 10^{-4}$, about three times larger.  The explanation for this is still unknown.  Furthermore, comparing $B(J/\psi \to \omega f_0(1710))\times B(f_0(1710)\to K^+K^-)$ with $B(J/\psi \to \omega f_0(1710))\times B(f_0(1710)\to \pi^+\pi^-)$ led to an upper limit on $B(f_0(1710)\to \pi\pi)/B(f_0(1710)\to K\bar{K})$ of $0.11$, apparently in contradiction with the finding from radiative decays.


These inconsistencies possibly point towards the need for more global analyses with higher statistics.  At BESIII, with 1.3~billion $J/\psi$ decays, this effort has just begun.  The $J/\psi \to \gamma \pi^0 \pi^0$ channel, for example, was recently reanalyzed with the full $J/\psi$ data set~\cite{Ablikim:2015umt}.  As a first step, rather than impose a resonant interpretation on the data, the $\pi^0 \pi^0$ mass spectrum was divided, bin-by-bin, into spin-0 and spin-2 components.  The spin-0 components are shown in the right panel of Figure~\ref{fig:JpsiRad}, and the shape is seen to be consistent with the BESII results. It is hoped that presenting the data in this way will encourage new ideas on how to parameterize the data.  These parameterizations can later be used to refit the data directly.

\subsection{S-wave $KK$, $\pi\pi$ and $K\pi$ scattering}

The details of S-wave $KK$, $\pi\pi$ and $K\pi$ scattering are beyond the scope of this review, but it should be mentioned that BESII has performed definitive work in this important area.  This has led to a more thorough understanding of the $f_0(980)$, the $\sigma$ and the $\kappa$.  

The $f_0(980)$ was seen prominently in the reaction $J/\psi \to \phi f_0(980)$, with the $f_0(980)$ decaying to both $\pi^+\pi^-$ and $K^+K^-$~\cite{Ablikim:2004wn}.  Since the mass of the $f_0(980)$ is close to the $K^+K^-$ threshold, its shape is distorted.  This fact can be used to study the coupling between the $\pi\pi$ and $K\bar{K}$ channels.  A simultaneous fit to the $f_0(980)$ in both decay modes was performed; the resulting coupling parameters are often still used today in experimental efforts to describe the $f_0(980)$.  

The $\sigma$ was studied in the channel $J/\psi \to \omega \pi^+ \pi^-$, where the $\sigma$ is seen in the $\pi^+\pi^-$ mass spectrum~\cite{Ablikim:2004qna}.  Again, the shape used to describe the $\sigma$ has had a major influence on many subsequent analyses.  

Finally, the $\kappa$ was studied in a very similar manner to the $\sigma$~\cite{Ablikim:2005ni}.  It is seen prominently in the $J/\psi \to K^*\bar{K}\pi + c.c.$ reaction, in the $\bar{K}\pi + c.c$ mass spectrum.  The cleanliness of this channel allowed a definitive study of the $\kappa$.

\subsection{Studies of the $X(1835)$}

The nature of the $X(1835)$ state (or states) remains one of the biggest mysteries in light quark physics at the BES experiments.  The first observation was at BESII in $J/\psi \to \gamma p\bar{p}$, where a large enhancement of events was seen around the $p\bar{p}$ threshold~\cite{Bai:2003sw}.  The enhancement was unexpected and was the source of much speculation.  This discovery paper remains the third most cited paper at BES.  The enhancement was confirmed at BESIII, first using $J/\psi$ decays coming from $\psi(2S)\to \pi^+\pi^- J/\psi$~\cite{BESIII:2010ad}, and then using 225~million directly-produced $J/\psi$~\cite{BESIII:2011aa}.  This latter analysis also measured the spin-parity of the enhancement to be $0^-$.

\begin{figure*}
\begin{center}
\includegraphics[width=1.0\linewidth]{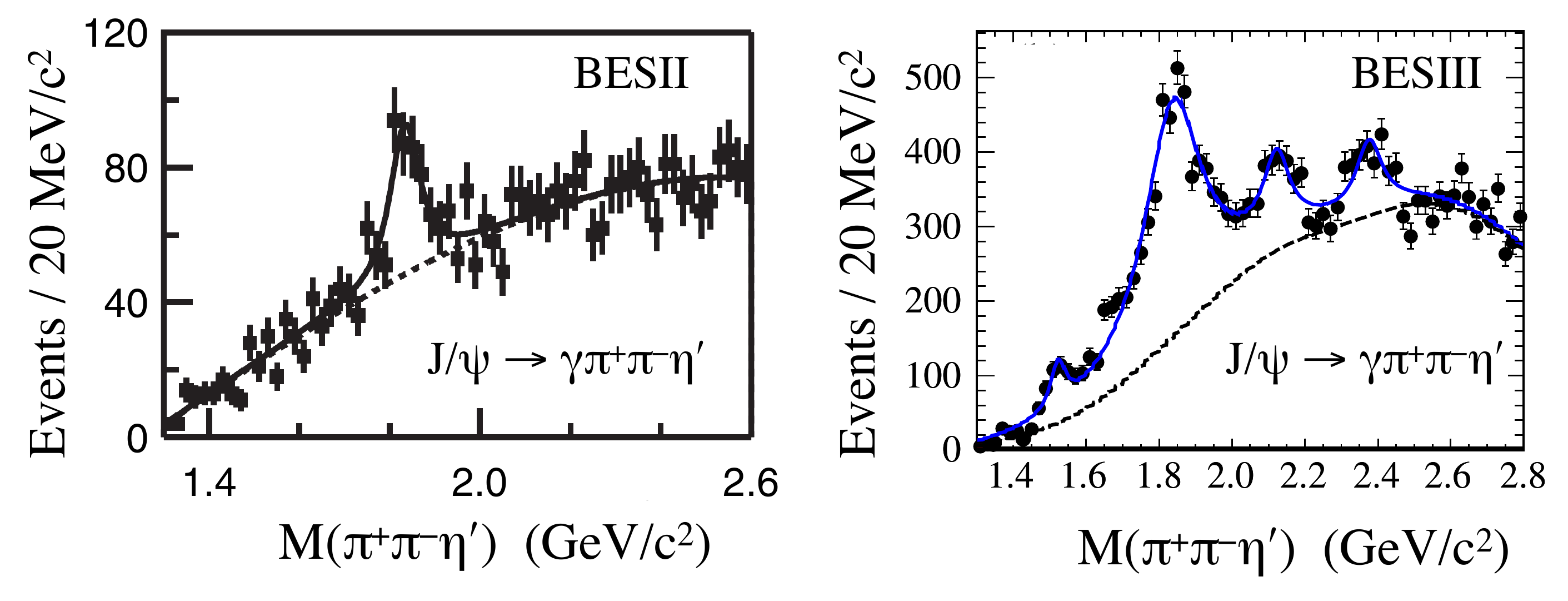}
\caption{Observation of the $X(1835)$ in $J/\psi \to \gamma X(1835); X(1835) \to \pi^+\pi^-\eta^\prime$ at BESII~\cite{Ablikim:2005um}~(left) and BESIII~\cite{BESIII:2011aa}~(right).  Additional states were discovered in the BESIII analysis.  Modified from References~\cite{Ablikim:2005um,BESIII:2011aa} with permission.}
\label{fig:X1835}
\end{center}
\end{figure*}

In parallel to the $p\bar{p}$ analyses, another peak at around the same mass and with about the same width was seen in $J/\psi \to \gamma \pi^+\pi^-\eta^\prime$ decays.  This was first seen by BESII~\cite{Ablikim:2005um}~(see the left panel of Figure~\ref{fig:X1835}).  BESIII was able to confirm the existence of this peak with increased statistics, but, surprisingly, also observed clear peaks at higher mass~\cite{Ablikim:2010au}~(see the right panel of Figure~\ref{fig:X1835}).  These high mass peaks are equally as mysterious as the $X(1835)$.    
In addition, another enhancement of events around 1.8~GeV/$c^2$ was seen in the related channel $J/\psi \to \gamma K_S^0 K_S^0 \eta$ at BESIII~\cite{Ablikim:2015toc}.  In this reaction, the enhancement was shown to have $J^P = 0^-$, the same as that for the $p\bar{p}$ enhancement.
It seems likely that the structures around 1.8~GeV/$c^2$ in $\pi^+\pi^-\eta^\prime$, $K_S^0 K_S^0 \eta$, and $p\overline{p}$ correspond to the same $X(1835)$, but there is as yet no definitive proof.  

A series of searches was also performed in other channels including a $p\bar{p}$ pair, such as $J/\psi \to \omega p \bar{p}$~\cite{Ablikim:2007ac}.  The lack of evidence for a $p\overline{p}$ threshold enhancement in these types of decays appears to disfavor a final-state interaction interpretation.  Other interpretations have been proposed, from a glueball state to a radial excitation of the $\eta$ to a baryonium state, but no definitive conclusion has yet been reached (see the references in~\cite{Ablikim:2015toc}).

\begin{marginnote}[0in]
  \entry{FINAL STATE INTERACTIONS}{Interactions among particles in the final state that can sometimes lead to misleading peaks not associated with resonances.}
\end{marginnote}

\begin{marginnote}[0.2in]
  \entry{BARYONIUM}{A bound state consisting of two baryons.}
\end{marginnote}

\subsection{Baryons in $J/\psi$ and $\psi(2S)$ decays}

In addition to the meson analyses described above, BES has also had a significant influence in light baryon spectroscopy.  Just as in the case of mesons, the well-defined initial state can be used to constrain properties of the final state.  For example, in the reaction $J/\psi \to p + X$, the $X$ baryon must have isospin-1/2 (in the absence of isospin violation), since the $J/\psi$ has isospin-0 and the proton has isospin-1/2.  It is thus a useful way to filter $N^*$ states from $\Delta$ states.  This type of filter is not available in fixed-target $\pi + N$ reactions, for example.

Studies of the $N^*$ states have been particularly fruitful, as can be seen in the following chain of analyses from BESI through BESIII.  In BESI, the $J/\psi \to p\bar{p}\eta$ channel was analyzed as a relatively simple one with which to begin~\cite{Bai:2001ua}.  The well-established $J^P = 1/2^+$ states $N(1535)$ and $N(1650)$ were clearly observed and their $J^P$ assignments were confirmed.  At BESII, this analysis was extended to the channel $J/\psi \to p\pi^-\bar{n} + c.c.$~\cite{Ablikim:2004ug}.  The same two states were observed, but in addition the $1/2^+$ $N(1440)$ was observed more clearly than other experiments (since it is usually eclipsed by the $\Delta$, which was absent in the BESII analysis).  And a new high-mass resonance, the $1/2^+$ or $3/2^+$ $N(2040)$ was found.  Finally, in BESIII, an analysis of $\psi(2S) \to p\bar{p}\pi^0$ was performed using 106~million $\psi(2S)$ decays~\cite{Ablikim:2012zk}.  Using the $\psi(2S)$ instead of the $J/\psi$ allowed an analysis of higher mass baryons and yet two more new ones were discovered, the $1/2^+$ $N(2300)$ and the $5/2^-$ $N(2570)$. Both of these states were observed with a significance of greater than 10$\sigma$.  These efforts from BES have been greatly influential in filling out the spectrum of $N^*$ states.

\begin{marginnote}[0.5in]
  \entry{N*'S AND $\Delta$'S}{Excited states of the proton and neutron.}
\end{marginnote}

The ability of BES to produce baryon resonances in $J/\psi$ and $\psi(2S)$ decays has made it a meaningful place to search for exotic baryons.  Such was the case in 2004, when there was much excitement about the pentaquark candidate $\Theta^+(1540)$.  BESII performed a search for this state in the $K_S^0 p K^- \bar{n}$ and $K_S^0 \bar{p} K^+ n$ decays of the $J/\psi$ and $\psi(2S)$~\cite{Bai:2004gk}.  The idea was that the pentaquark might be produced in pairs (to conserve flavor and other quantum numbers).  No evidence was found for the pentaquark decaying to $K_S^0 p$ or $K^+ n$ and tight upper limits were placed on its production.  Eventually, the initial evidence for the pentaquark was overturned.  The BESII search was among the earliest of the negative searches.


\section{CHARMONIUM PHYSICS}

\label{sec:charmonium}

While the study of light quark physics is generally associated with the $J/\psi$ data sets at BES, the study of charmonium is most often pursued through data taken at the $\psi(2S)$. 
From the $\psi(2S)$, all charmonium states below $D\bar{D}$ threshold can be reached, making the $\psi(2S)$ data ideal for charmonium studies.
The $\chi_{cJ}(1P)$ can be accessed through E1 radiative transitions; the $\eta_c(1S,2S)$ through M1 radiative transitions; and the $J/\psi$ and $h_c(1P)$ states through hadronic transitions.
BESI, BESII, and BESIII have all collected increasingly large samples of $\psi(2S)$ decays, and there have been important results from each.  BESI collected 3.8~million $\psi(2S)$ decays; BESII collected 14~million; and BESIII took an initial sample of 106~million in 2009 and increased it to 448~million in 2012.

One of the most interesting features of the states below $D\bar{D}$ threshold is that they can be successfully described by treating the $c\bar{c}$ pair as being bound in a potential.  Studying the masses and radiative transitions of the charmonium states gives valuable insight into the shape of the potential.  
On the other hand,
the shape of the potential and its spin-dependence can be derived from QCD (lattice QCD) or phenomenologically.
Furthermore, masses and radiative transitions can now be directly calculated in lattice QCD.
The properties of charmonium thus provide a convenient point of contact between experiment and QCD.  See Ref.~\cite{Brambilla:2010cs} for a review of issues in charmonium.

\begin{marginnote}[0in]
\entry{Lattice QCD (LQCD)}
{A method for calculating strong interaction quantities on computers, 
with space-time represented by a discrete lattice.  
}
\end{marginnote}

The following two sections will cover a selection of BES results on masses and radiative transitions of charmonium states below $D\bar{D}$ threshold. The final section will discuss some anomalies in $J/\psi$ and $\psi(2S)$ decays.

\begin{textbox}[h]
\section{Charmonium}
A charmonium state is made of a charm quark and an anti-charm quark with a given set of internal quantum numbers, such as spin~($S$), orbital angular momentum~($L$), and principal quantum number~($n$). The charmonium system is the set of all possible charmonium states. It is thus similar to the Hydrogen atom or the positronium system in Quantum Electrodynamics.    Unlike Hydrogen or positronium, however, each state of charmonium has a different name.  The $\eta_c(1S)$ is the ground state, with $n=1$, $L=0$, and $S=0$.  The $h_c(1P)$, as another example, has $n=1$, $L=1$, and $S=0$.  
\end{textbox}

\subsection{Masses of charmonium states}

As mentioned above, masses of charmonium states provide key information about the form of the potential binding the $c\bar{c}$ pair.  The $\eta_c(1S)$ plays a special role since it is the ground state of charmonium. 
Furthermore, 
since the $\eta_c(1S)$ and $J/\psi$ only differ in their spin (the $\eta_c(1S)$ has $S=0$ while the $J/\psi$ has $S=1$), their mass difference, also known as the hyperfine splitting, is sensitive to the 
spin-spin part of the potential.
The calculation of the hyperfine splitting is a key prediction of many models.
The mass splitting between the $h_c(1P)$ and the $\chi_{cJ}(1P)$ states (combined in the form of a spin-weighted average) plays a similar role.
These states also differ only in their spin (the $h_c(1P)$ has $S=0$, while the $\chi_{cJ}(1P)$ has $S=1$), but their internal orbital angular momentum ($L$) is one.
In this case it is expected that the mass splitting vanishes to lowest order.
Thus a measurement of the mass splitting is sensitive to higher order effects.

BES has made important and unique contributions to the measurement of the masses of the $\eta_c(1S)$, $h_c(1P)$, and $\chi_{cJ}(1P)$.  The measurement of each is summarized below.

\begin{figure*}
\begin{center}
\includegraphics[width=1.0\linewidth]{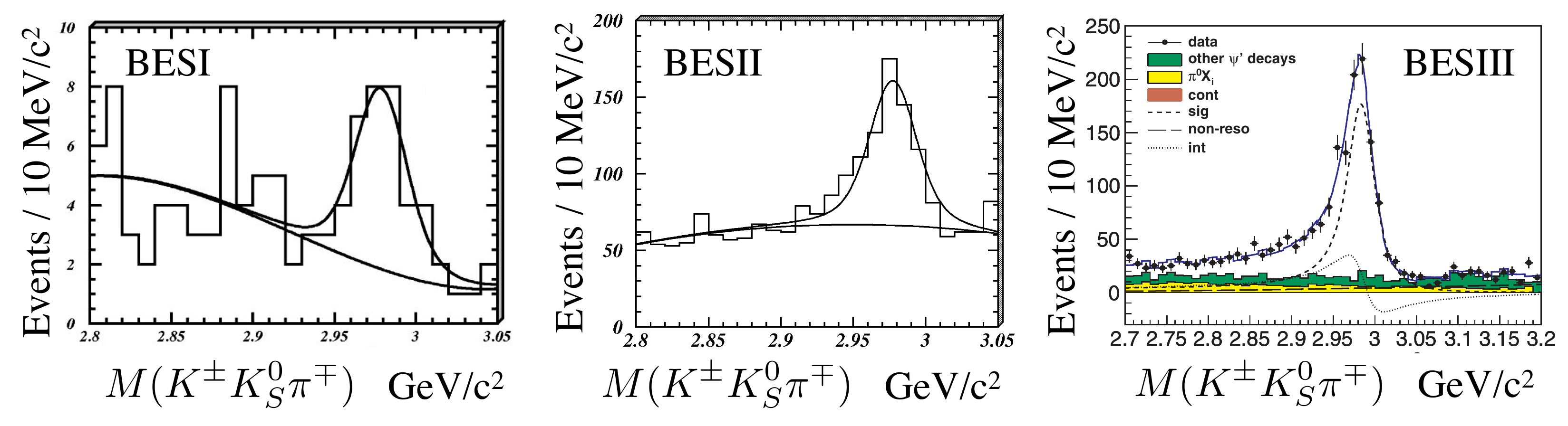}
\caption{Evolution of measurements of the $\eta_c(1S)$ mass. 
(left) Measurement of the $\eta_c(1S)$ mass at BESI~\cite{Bai:2000sr} 
in the process $J/\psi\to \gamma \eta_c(1S); \eta_c(1S) \to K^{\pm}K_S^0\pi^\mp$
using 7.8~million $J/\psi$ decays.
(middle) Measurement of the $\eta_c(1S)$ mass at BESII~\cite{Bai:2003et} in the same process
using 58~million $J/\psi$ decays.
(right) Measurement of the $\eta_c(1S)$ mass at BESIII~\cite{BESIII:2011ab} in the same process except from $\psi(2S)$
using 106~million $\psi(2S)$ decays.
In each case the $K^{\pm}K_S^0\pi^\mp$ decay mode of the $\eta_c(1S)$ is shown as an example -- each analysis used a combination of a number of different $\eta_c(1S)$ decays.
In both BESI and BESII~(the left two plots), the $\eta_c(1S)$ peak was fit with a symmetric Breit-Wigner distribution.  In BESIII~(the right plot), an $E_\gamma^7$ term was added, and interference with the non-$\eta_c(1S)$ background was allowed.  Notice the obvious distortion in the lineshape at BESIII and hints of the same distortion at BESII.  Modified from References~\cite{Bai:2000sr,Bai:2003et,BESIII:2011ab} with permission.
}
\label{fig:etac}
\end{center}
\end{figure*}

{\bf\boldmath 1. Measurement of the mass of the $\eta_c(1S)$.}
BES has a long history of $\eta_c(1S)$ mass measurements using the M1 transitions $J/\psi \to \gamma \eta_c(1S)$ and $\psi(2S) \to \gamma \eta_c(1S)$.  BESI, combining results from 7.8~million $J/\psi$ and 3.8~million $\psi(2S)$ decays found the mass to be 
$2976.3 \pm 2.3 \pm 1.2$~MeV/$c^2$~\cite{Bai:1998cw,Bai:2000sr}~(see the left panel of Figure~\ref{fig:etac});
and BESII, using 58~million $J/\psi$ decays, found 
$2977.5 \pm 1.0 \pm 1.2$~MeV/$c^2$~\cite{Bai:2003et}~(see the middle panel of Figure~\ref{fig:etac}).
These measurements were consistent with other measurements from radiative decays,
but systematically lower than measurements from other production mechanisms, such as $\gamma\gamma$ collisions, $B$ decays, or $p\bar{p}$ annihilation.
This represented a serious problem.
At BESIII, using 106~million $\psi(2S)$ decays, the line-shape of the $\eta_c(1S)$ was found to be clearly distorted~(see the right panel of Figure~\ref{fig:etac}).  Taking into account the expected $E^7$ energy-dependence of the radiated photon~\cite{Mitchell:2008aa}, and including interference with the non-$\eta_c(1S)$ background, the mass was found to be 
$2984.3 \pm 0.6 \pm 0.6$~MeV/$c^2$~\cite{BESIII:2011ab}, more in line with other measurements, and resolving the previous discrepancy.  The lower statistics of the previous measurements apparently hid these important effects.  A subsequent measurement using $h_c(1P)\to\gamma\eta_c(1S)$ confirmed this higher mass~\cite{Ablikim:2012ur}.

{\bf\boldmath 2. Measurement of the mass of the $\chi_{cJ}(1P)$.}
BESII measured the masses of the $\chi_{cJ}(1P)$ states using the process $\psi(2S) \to \gamma \chi_{cJ}(1P)$~\cite{Ablikim:2005yd}.  
The $\chi_{cJ}(1P)$ were allowed to decay inclusively.
Rather than detect the energy of the photon directly, events in which the photon converted in the detector to an $e^+e^-$ pair were used.
This allowed a much better determination of the photon energy, with resolutions on the order of 2-4~MeV.  The spin-weighted average mass of the $\chi_{cJ}(1P)$ was determined to be $3524.85 \pm 0.32 \pm 0.30$~MeV/$c^2$.  Despite being over a decade old, this measurement still represents a major component of the world average.  It is surpassed in precision only by measurements in $p\bar{p}$ annihilation~\cite{Andreotti:2005ts}.

{\bf\boldmath 3. Measurement of the mass of the $h_c(1P)$.}
BESIII has made two measurements of the $h_c(1P)$ mass, both using the transition $\psi(2S)\to\pi^0 h_c(1P)$.  In the first, the $h_c(1P)$ was reconstructed both inclusively and by tagging the photon in the transition $h_c(1P)\to\gamma \eta_c(1S)$~\cite{Ablikim:2010rc}.  Even though the inclusive process has a large background, the measurement of the mass had a total error (statistical and systematic combined) of around 200~keV.  This analysis will be discussed further below in the context of the measurement of $B(h_c(1P)\to \gamma \eta_c(1S))$.  The second measurement, however, was even more precise.  In this analysis, the process $\psi(2S)\to\pi^0 h_c(1P); h_c(1P)\to\gamma \eta_c(1S)$ was reconstructed exclusively using 16 decay modes of the $\eta_c(1S)$~\cite{Ablikim:2012ur} .  This allowed an extremely clean sample of over 800 $h_c(1P)$ events.  The mass was determined to be $3525.31 \pm 0.11 \pm 0.14$~MeV/$c^2$ and the width was measured as $0.70 \pm 0.28 \pm 0.22$~MeV/$c^2$.  Both are the most precise measurements to date.



\subsection{Radiative transitions between charmonium states}

BES has also made a number of influential measurements of 
radiative transitions among charmonium states.
A few of its unique contributions are highlighted below.

\begin{figure*}
\begin{center}
\includegraphics[width=1.0\linewidth]{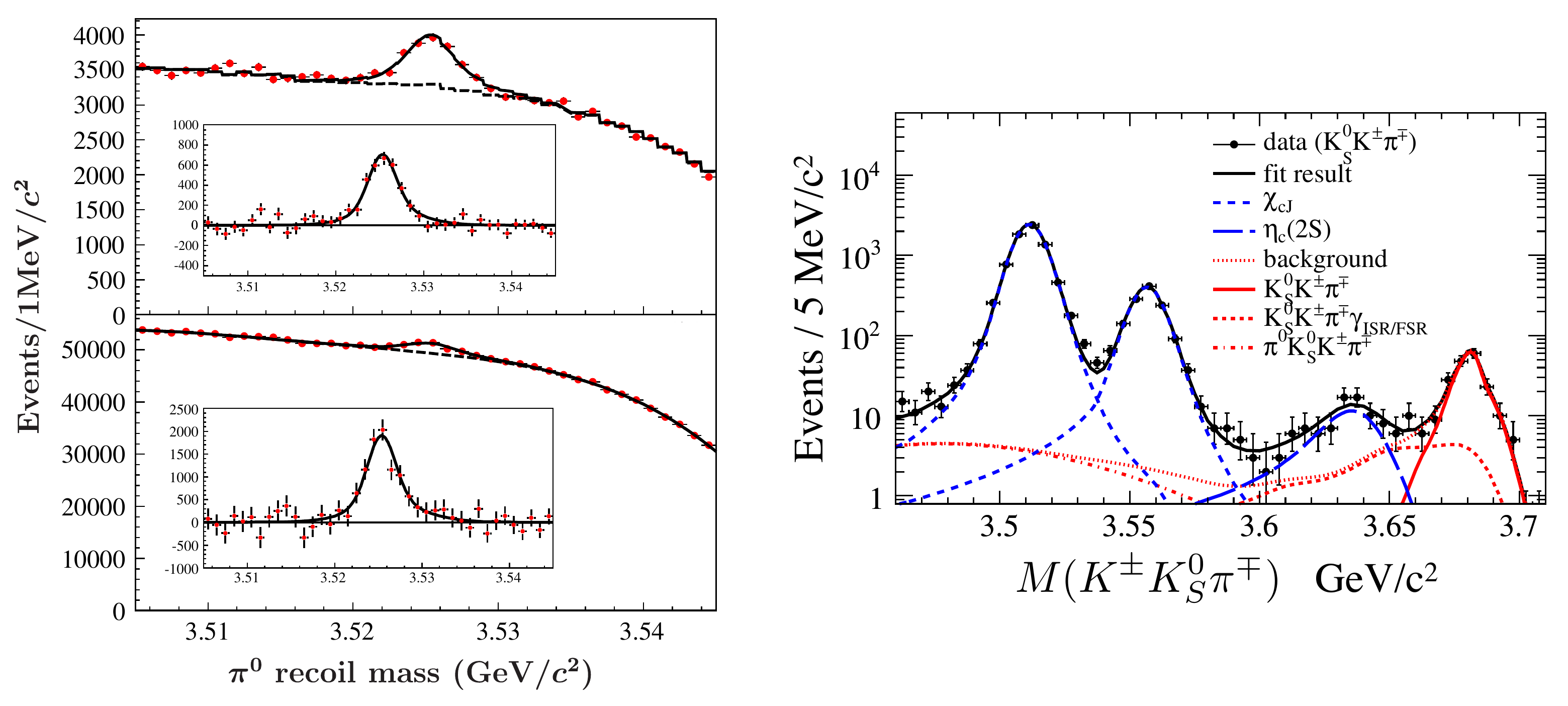}
\caption{Measurements of radiative transitions at BESIII.  (left)~The top plot shows a fit to the $h_c(1P)$ mass in the process $\psi(2S)\to \pi^0 h_c(1P); h_c(1P)\to\gamma\eta_c(1S)$.  The energy of the radiated photon is used to tag this process.  The bottom plot is a fit to the $h_c(1P)$ mass in the process $\psi(2S)\to \pi^0 h_c(1P)$.  The ratio of the two fits gives $B(h_c(1P) \to \gamma \eta_c(1S))$~\cite{Ablikim:2010rc}. (right)~The measurement of $B(\psi(2S)\to\gamma\eta_c(2S))$~\cite{Ablikim:2012sf}.  The $\eta_c(2S)$ is the peak between 3.60 and 3.65~GeV/$c^2$, surrounded by prominent backgrounds on either side.  Modified from References~\cite{Ablikim:2010rc,Ablikim:2012sf} with permission.
}
\label{fig:transitions}
\end{center}
\end{figure*}

{\bf\boldmath 1. Measurement of $B(h_c(1P) \to \gamma \eta_c(1S))$. }
Using its initial sample of 106~million $\psi(2S)$ decays, BESIII was able to make the first measurement of the E1 transition rate $B(h_c(1P) \to \gamma \eta_c(1S))$~\cite{Ablikim:2010rc}.
This was measured by fitting the $h_c(1P)$ peak in the inclusive $\pi^0$ recoil mass spectrum with and without tagging the photon from $h_c(1P) \to \gamma \eta_c(1S)$.  When the photon is tagged, the fit gives the product $B(\psi(2S)\to \pi^0 h_c(1P))\times B(h_c(1P) \to \gamma \eta_c(1S))$~(see the top histogram in the left panel of Figure~\ref{fig:transitions}).  When the photon is not tagged, the fit gives $B(\psi(2S)\to \pi^0 h_c(1P))$~(see the bottom histogram in the left panel of Figure~\ref{fig:transitions}).  Dividing these two results gives a measurement of $B(h_c(1P) \to \gamma \eta_c(1S)) = (54.3 \pm 6.7 \pm 5.2)\%$.  This measurement falls within the wide range of expected values, and has helped restrict theoretical models.

{\bf\boldmath 2. Measurement of $B(\psi(2S) \to \gamma \eta_c(2S))$. }
Unsuccessful searches for the transition $\psi(2S) \to \gamma \eta_c(2S)$ have been carried out since the early 1980's.  BESIII was finally able to make the first observation of this process using its initial sample of 106~million $\psi(2S)$ decays~\cite{Ablikim:2012sf}.  The low energy of the transition photon and the prominent background peaks due to $\psi(2S) \to \gamma \chi_{cJ}(1P)$  make this an especially difficult measurement.  To reduce background, the $\eta_c(2S)$ was reconstructed in the exclusive channels $K_S^0 K^\pm \pi^\mp$ and $K^+K^-\pi^0$.  The resulting $K_S^0 K^\pm \pi^\mp$ mass spectrum is shown in the right panel of Figure~\ref{fig:transitions} (the $K^+K^-\pi^0$ mass spectrum is similar, but not shown).  The signal appears as the peak between $3.60$ and $3.65$~GeV/$c^2$.  The peaks to the left of the signal are from the $\chi_{cJ}$; the peak to the right is from spurious showers in the calorimeter.  Normalizing to a BaBar measurement of $B(\eta_c(2S) \to K\bar{K}\pi)$~\cite{Aubert:2008kp} gives a branching fraction of $B(\psi(2S)\to \gamma \eta_c(2S)) = (6.8 \pm 1.1 \pm 4.5)\times 10^{-4}$.  This remains the only observation of this process.

\begin{marginnote}[0.5in]
  \entry{MULTIPOLES}{Terms in an expansion of the radiative transition amplitude.  Higher-order terms are suppressed, but are sensitive to details of the transition.}
\end{marginnote}

{\bf\boldmath 3. Measurement of multipoles in $\psi(2S) \to \gamma \chi_{c2}$. }
The high statistics and cleanliness of the process $\psi(2S) \to \gamma \chi_{c2}; \chi_{c2} \to \pi^+\pi^-$ and $K^+K^-$ have allowed detailed studies of multipoles beyond the dominant E1 transition.  These higher multipoles are important for a number of reasons:  they could serve as an explanation for some apparent deviations from theoretical E1 rates;
 the M2 amplitude is sensitive to the anomalous magnetic moment of the charm quark;
 and the E3 amplitude is sensitive to the orbital angular momentum of the quarks in the $\psi(2S)$.  An initial measurement from BESII~\cite{Ablikim:2004qn}, using 14~million $\psi(2S)$ decays, found M2 and E3 contributions consistent with zero.  The measurement from BESIII~\cite{Ablikim:2011da}, using 106~million $\psi(2S)$, gave the first evidence of a non-zero M2 component.  It is inconsistent with zero with a significance of $4.4\sigma$.  It is consistent with predictions when the anomalous magnetic moment of the charm quark is assumed to be zero.  An improved result from BESIII, using the full 448~million $\psi(2S)$ decays dataset, is forthcoming.

\subsection{Decays of the $J/\psi$ and $\psi(2S)$}

Another interesting feature of charmonium physics is the surprising differences between $J/\psi$ and $\psi(2S)$ decays to light quark states.
It is reasonable to think that once the charm and anti-charm quarks of the initial $J/\psi$ or $\psi(2S)$ annihilate, predominantly going through a single virtual photon or three gluons, the subsequent hadronization of the photon or gluons should be independent of their origin.
From this reasoning, one would expect that the ratio of rates for the $\psi(2S)$ and $J/\psi$ to decay to any specific combination of light quark hadrons would be roughly constant (after adjusting for the mass difference of the $J/\psi$ and $\psi(2S)$ in a straightforward way).
Since the rate for the dilepton decay of $\psi(2S)$ is roughly 12\% that of the $J/\psi$, it is thought that this constant ratio should be around 12\%.
This is the ``12\% rule.''

The 12\% rule does, in fact, hold for many decays of the $J/\psi$ and $\psi(2S)$. For example, BESII made the best measurement of the branching fraction $B(J/\psi \to p\bar{p}\pi^0)$~\cite{Ablikim:2009iw}, while BESIII made the best measurement of $B(\psi(2S) \to p\bar{p}\pi^0)$~\cite{Ablikim:2012zk}.  
Taking the ratio of the world average values (dominated by the BES measurements)
one finds $B(\psi(2S) \to p\bar{p}\pi^0)/B(J/\psi \to p\bar{p}\pi^0) = (12.9 \pm 1.0)\%$,
consistent with the 12\% rule.


However, the 12\% rule fails spectacularly for a few decay channels.
One of the most well-known is in $J/\psi$ and $\psi(2S)$ decays to $\rho\pi$, where the $\rho\pi$ decays to the $\pi^+\pi^-\pi^0$ final state~\cite{Franklin:1983ve}.
Using a combination of 58~million directly produced $J/\psi$ and $J/\psi$ produced using 14~million $\psi(2S)$, BESII determined
$B(J/\psi\to\rho\pi) = (2.10 \pm 0.12) \times 10^{-2}$~\cite{Bai:2004jn}.
In contrast, BESII performed a PWA of the $\psi(2S)\to\pi^+\pi^-\pi^0$ channel to determine 
$B(\psi(2S)\to\rho\pi) = (5.1 \pm 1.3) \times 10^{-5}$~\cite{Ablikim:2005jy}.
The ratio of $\psi(2S)$ to $J/\psi$ is only $(0.24 \pm 0.06)\%$, much smaller than 12\%.
This phenomenon is referred to as the $\rho\pi$ puzzle; a definitive solution is yet to be found.
In addition to the vastly different rates of $\rho\pi$ production in $\psi(2S)$ and $J/\psi$ decays, there is also a striking difference between their $\pi^+\pi^-\pi^0$ Dalitz plots.  BESIII published a stark comparison in Ref.~\cite{Ablikim:2012dx}.

Two other interesting decays where the 12\% rule fails are $\gamma \eta$ and $\gamma \eta^\prime$.  BESI did an early analysis of the $\psi(2S)$ decays~\cite{Bai:1998ny}; and
BESII did an early analysis of $J/\psi$ decays~\cite{Ablikim:2005je}.
BESIII made a definitive measurement of the $\psi(2S)$ decays~\cite{Ablikim:2010dx}, the most precise measurements to date.
The ratio of branching fractions to $\gamma\eta^\prime$ is $(2.4 \pm 0.1)\%$, violating the 12\% rule.
But even more dramatic is the $\gamma\eta$ channel, where the ratio of $\psi(2S)$ to $J/\psi$ decays is only 
$(0.13 \pm 0.05)\%$,
even lower than the $\rho\pi$ ratio.
In addition to violating the 12\% rule, it is also surprising that the ratios for $\gamma\eta$ and $\gamma\eta^\prime$ are so different from one another.


\section{XYZ PHYSICS}

\label{sec:XYZ}

Apart from a few anomalies, such as the $\rho\pi$ puzzle, discussed above, the charmonium system below $D\bar{D}$ threshold is fairly well understood. 
The same is not true for the states above $D\bar{D}$ threshold.  
Starting with the discoveries of the $X(3872)$ in 2003 at Belle~\cite{Choi:2003ue} and the $Y(4260)$ in 2005 at BaBar~\cite{Aubert:2005rm}, there has been a flood of new states that cannot be accommodated within the $c\bar{c}$ picture of charmonium.  
These anomalous states, referred to as the ``$XYZ$'' states (reflecting their still-mysterious nature), could be pointing towards the existence of exotic compositions of quarks and gluons~\cite{Brambilla:2010cs}.  

For example, the $Y(4260)$ could be a ``hybrid meson," a meson made of a quark and an anti-quark (as in a ``conventional meson''), but with the gluonic field in an excited state.  The $X(3872)$, on the other hand, could be a ``meson molecule,'' a meson composed of a bound state of two conventional mesons.  Other possibilities for the $XYZ$ are ``tetraquarks'' (composites of two quarks and two anti-quarks),
 or ``hadrocharomium'' (conventional mesons surrounded by a field of light quark mesons), among others~\cite{Brambilla:2010cs}.  

The existence of non-$q\bar{q}$ states would help clarify our understanding of QCD, which, according to the latest calculations, predicts them.
It is also possible that a few of these observed phenomena may not actually be ``states'' at all, but instead arise from rescattering effects, or the opening of thresholds, etc.  
If this turns out to be the case, then the $XYZ$ region would provide a prime testing ground for understanding such phenomena.  In any case, studies of the $XYZ$ are continually breaking new ground, and the issues that have arisen have not yet been resolved.

The ``Y'' family of states is especially relevant for the BESIII studies that will be discussed below.  They are produced in the process $e^+e^- \to Y$, where the center-of-mass collision energy of the $e^+e^-$ matches the mass of the produced $Y$. But before BESIII, they were studied primarily at Belle and BaBar, where the center-of-mass energies of the $e^+e^-$ collisions are typically in the 10~GeV region, far above the masses of the $Y$ states, which are in the region of $4-5$~GeV/$c^2$.  To produce them, Belle and BaBar relied on initial state radiation~(ISR), a relatively rare process whereby the initial $e^+$ or $e^-$ first radiates a high-energy photon before annihilating, reducing the center-of-mass collision energy to the required region.  

The breakthrough at BESIII was to produce these states directly, taking advantage of the more propitious energy range of BEPC.  Thus, the $Y(4260)$ could be produced by tuning the $e^+e^-$ center-of-mass energy to 4.26~GeV, the $Y(4360)$ could be produced at 4.36~GeV, and so on.  This has at least two advantages.  The rates are higher, because the process does not depend on the emission of a high-energy ISR photon.  Also, the $Y$ is produced at rest in the laboratory, as opposed to boosted along the beam direction as in the ISR process, making the detection of the final decay products more efficient.

The initial idea at BESIII was to collect 500~pb$^{-1}$ of data at both 4.26 and 4.36~GeV in 2013 in order to study decays of the $Y(4260)$ and $Y(4360)$, respectively.  However, after many discoveries, such as the discoveries of the charged $Z_c$ states, to be discussed below, the program was extended.  After an extended running period in 2013 and another year of running in 2014, BESIII now has large samples of events at 4.23~GeV~(1092~pb$^{-1}$), 4.26~GeV~(826~pb$^{-1}$), 4.36~GeV~(540~pb$^{-1}$), 4.42~GeV~(1074~pb$^{-1}$), and 4.60~GeV~(567~pb$^{-1}$), as well as smaller samples at many energy points between~\cite{Ablikim:2015nan}.  This is in addition to the 482~pb$^{-1}$ of data collected at 4.01~GeV in 2011.

\subsection{Discovery of charged $Z_c$ states}

The initial samples of 500~pb$^{-1}$ of $e^+e^-$ collision data at 4.26 and 4.36~GeV were collected between mid-December of 2012 and February of 2013.
One of the first channels to be checked, even before data-taking had finished, was $e^+e^- \to \pi^+\pi^-J/\psi$ at 4.26~GeV, since this is near the peak of the $Y(4260)$, and the $Y(4260)$ is known to decay to $\pi^+\pi^-J/\psi$.
Initial checks of the cross section agreed with what was expected based on the Belle and BaBar measurements of the same channel using the ISR process.
But it was also quickly noticed that there was a large peak around 3900~MeV/$c^2$ in the $\pi^\pm J/\psi$ subsystem.
Such a peak, subsequently named the $Z_c(3900)$, points towards the existence of a particle that is manifestly exotic.
Decaying to the $J/\psi$, it most likely contains a $c\bar{c}$ pair.  But being charged, it must include more than that $c\bar{c}$ pair.
The simplest interpretation is that its electric charge comes from an additional light quark and anti-quark pair, making it a strong candidate for a tetraquark or a meson molecule, among other possibilities.

\begin{figure*}
\begin{center}
\includegraphics[width=1.0\linewidth]{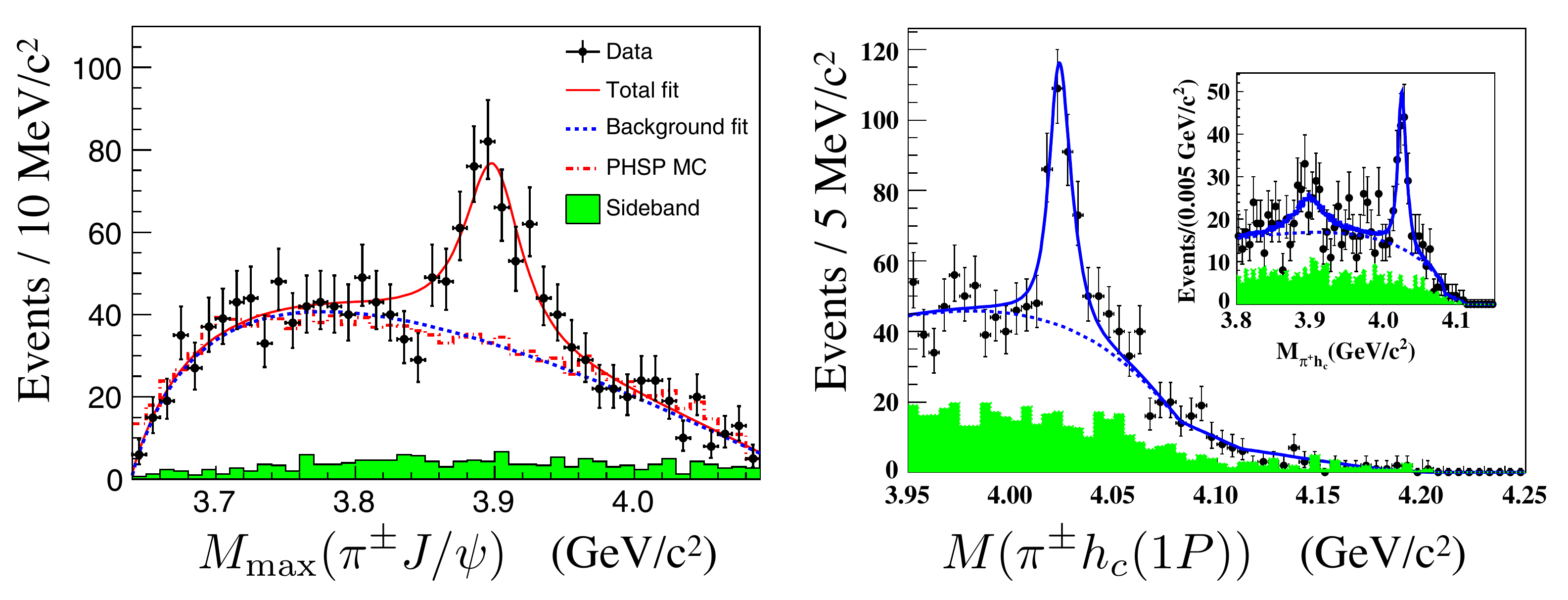}
\caption{Discoveries of the $Z_c(3900)$ and $Z_c(4020)$ at BESIII.  (left) Discovery of the $Z_c(3900)$ in the $\pi^\pm J/\psi$ substructure of the $e^+e^-\to\pi^+\pi^-J/\psi$ reaction~\cite{Ablikim:2013mio}.  (right) Discovery of the $Z_c(4020)$ in the $\pi^\pm h_c(1P)$ substructure of $e^+e^-\to\pi^+\pi^-h_c(1P)$~\cite{Ablikim:2013wzq}.
The points are data and the solid~(green) histogram shows the background estimate from the $J/\psi$~(left) and $h_c(1P)$~(right) sidebands.  The inset in the right plot shows a search for the $Z_c(3900)$ decaying to $\pi^\pm h_c(1P)$.    Modified from References~\cite{Ablikim:2013mio,Ablikim:2013wzq} with permission.
}
\label{fig:Zc}
\end{center}
\end{figure*}

The analysis of the $e^+e^- \to \pi^+\pi^-J/\psi$ process at 4.26~GeV, and the observed charged $Z_c(3900)$ in the $\pi^\pm J/\psi$ subsystem, was performed quickly, but with many cross-checks.  
The result was made public in March of 2013 and was published in June~\cite{Ablikim:2013mio}, only four months after the data was taken.
A simultaneous observation of the $Z_c(3900)$, but with fewer events, was published by Belle~\cite{Liu:2013dau}.  In fact, a few of the primary authors on the BESIII paper were also among the primary authors of the Belle paper.
The $Z_c(3900)$, as seen by BESIII, is shown in the left panel of Figure~\ref{fig:Zc}.
Its mass and width were found to be
$3899.0 \pm 3.6 \pm 4.9$~GeV/$c^2$ and
$46 \pm 10 \pm 20$~GeV/$c^2$, respectively.
Although only published in 2013, the observation of the $Z_c(3900)$ is already the highest cited paper at BESIII, having received over 300~citations.

In February of 2013, shortly after the end of the initial round of data-taking, there was a BESIII collaboration meeting at Tsinghua University.  Many surprising results from the new data sets were shown (some of which are discussed below), and it was decided to extend the data-taking time until June of 2013. 


The first of these additional surprises was the discovery of the charged $Z_c(4020)$ that appears in the $\pi^\pm h_c(1P)$ subsystem of the process $e^+e^- \to \pi^+\pi^-h_c(1P)$~\cite{Ablikim:2013wzq}.
The discovery is shown in the right panel of Figure~\ref{fig:Zc}.  Its mass and width were determined to be 
$4022.9 \pm 0.8 \pm 2.7$~GeV/$c^2$ and
$7.9 \pm 2.7 \pm 2.6$~GeV/$c^2$, respectively.
The reasons that this state is interesting  are the same as those for the $Z_c(3900)$:  it decays to charmonium and it is charged.  It is therefore an additional tetraquark (or meson molecule) candidate.

One clue about the nature of the $Z_c(3900)$ and the $Z_c(4020)$ may come from their masses.  
The $Z_c(3900)$ has a mass just above $D^*\bar{D}$ threshold and the $Z_c(4020)$ has a mass just above $D^*\bar{D}^*$ threshold.
Thus, along with the closed charm channels, analyses were simultaneously performed in the open charm reactions $e^+e^- \to (D\bar{D}^*)^\pm \pi^\mp$ and $(D^*\bar{D}^*)^\pm \pi^\mp$.
In each case, a peak was found just above the charged $D^*\bar{D}^{(*)}$ threshold.  In the case of the  $(D\bar{D}^*)^\pm \pi^\mp$ channel, the peak was measured to have a mass and width of 
$3883.9 \pm 1.5 \pm 4.2$~GeV/$c^2$ and
$24.8 \pm 3.3 \pm 11.0$~GeV/$c^2$, respectively~\cite{Ablikim:2013xfr} .  By looking at the angular distribution of its decay, this peak was conclusively found to have $J^P = 1^+$.  This result was also confirmed using a more exclusive method of reconstruction~\cite{Ablikim:2015swa}.
And in the case of the $(D^*\bar{D}^*)^\pm \pi^\mp$ channel, the peak occurred at a mass of $4026.3 \pm 2.6 \pm 3.7$~GeV/$c^2$ and had a width of $24.8 \pm 5.6 \pm 7.7$~GeV/$c^2$.~\cite{Ablikim:2013emm}.
While the masses and widths of the open and closed charm peaks are slightly different, it is reasonable to assume these phenomena are related.

\subsection{Emerging patterns and problems}

One of the goals of the BESIII $XYZ$ physics program is to establish patterns among the multitude of new states.  For example, it seems possible the interpretation of the $Y(4260)$ is somehow related to the interpretation of the $Z_c(3900)$ and $Z_c(4020)$, since the latter are possibly produced in the decays of the former.  To establish this hypothesis, though, the $e^+e^- \to Z_c(3900,4020)^\pm \pi^\mp$ cross sections need to be mapped as a function of $e^+e^-$ center-of-mass energy to see if they follow the shape of the $Y(4260)$.

Another connection between the $XYZ$ states was also possibly found through the observation of the process $e^+e^- \to \gamma X(3872)$~\cite{Ablikim:2013dyn}.  Mapping the cross section as a function of $e^+e^-$ center-of-mass energy does appear to trace out the $Y(4260)$.  However, more data is needed to show this conclusively.  It is hoped that connections such as these among established $XYZ$ states will aid in their interpretation.

Another satisfying set of results was the observation of neutral partners to the charged $Z_c(3900)$ and $Z_c(4020)$.  In this series of analyses, the neutral partner to the $Z_c(3900)$ was seen in 
the $\pi^0 J/\psi$ subsystem 
of $e^+e^- \to \pi^0\pi^0 J/\psi$~\cite{Ablikim:2014dxl}; 
the neutral partner to the $Z_c(4020)$ was seen in 
the $\pi^0 h_c(1P)$ subsystem 
of $e^+e^- \to \pi^0\pi^0 h_c(1P)$~\cite{Ablikim:2015tbp}; 
the neutral partner to the charged $D\bar{D}^*$ state 
(presumably related to the $Z_c(3900)$) was found in 
the neutral $D\bar{D}^*$ subsystem 
of $e^+e^- \to \pi^0(D\bar{D}^*)^0$~\cite{Ablikim:2015gda};
and the neutral partner to the charged $D^*\bar{D}^*$ state 
(presumably related to the $Z_c(4020)$) was found in 
the neutral $D^*\bar{D}^*$ subsystem 
of $e^+e^- \to \pi^0(D^*\bar{D}^*)^0$~\cite{Ablikim:2015vvn}.

Finally, connections are also possibly emerging between the charmonium and strangeonium systems.  BESII observed a state called the $Y(2175)$ (originally observed by BaBar using ISR~\cite{Aubert:2006bu}) in the decay $J/\psi \to \eta Y(2175)$ with $Y(2175) \to f_0(980) \phi$ and $f_0(980) \to \pi^+\pi^-$~\cite{Ablikim:2007ab}.  It was confirmed with higher statistics at BESIII~\cite{Ablikim:2014pfc}.  This state is thought to possibly be the strangeonium analog of the $Y(4260)$.

But alongside the emergence of these patterns have come new problems.  The most prominent of these is the behavior of exclusive $e^+e^-$ cross sections as a function of center-of-mass energy, where there currently appears to be little order.  It was previously known that the $Y(4260)$ appears in the $e^+e^- \to \pi^+\pi^- J/\psi$ cross section, but does not appear in the $e^+e^- \to \pi^+\pi^- \psi(2S)$ cross section.  Instead, $e^+e^- \to \pi^+\pi^- \psi(2S)$ shows two clear structures, one called the $Y(4360)$ and one called the $Y(4660)$.  BESIII has already added to this mystery by measuring a number of other channels.  The $e^+e^- \to\pi^+\pi^- h_c(1P)$ cross section~\cite{Ablikim:2013wzq} shows no evidence for the $Y(4260)$, the $Y(4360)$, or the $Y(4660)$, only a very broad hump and possibly a narrow peak around 4.23~GeV/$c^2$.  The $\eta J/\psi$ cross section~\cite{Ablikim:2012ht,Ablikim:2015xhk} is also inconsistent with $\pi^+\pi^- J/\psi$, but a finer scan is needed to determine the energy-dependence.  Finally, the $\omega \chi_{c0}$ cross section~\cite{Ablikim:2014qwy} was seen to peak near threshold, then quickly fade away.

What are the mechanisms that cause the cross sections to behave so differently?  And why, in general, is so much closed charm being produced so far above open charm thresholds?
With more data in the coming years, BESIII will be capable of adding valuable information concerning these issues.  And there will likely be more surprises.


\section{CHARM PHYSICS}

At colliders operating near charm threshold, studies of the physics 
of $D^0$ and  $D^+$ mesons are performed primarily 
via data taken at the $\psi(3770)$ resonance.  
This is third-lowest $J^{PC} = 1^{--}$ state (the quantum numbers 
directly accessible in $e^+e^-$ collisions)
and the first with a mass above the $D\bar{D}$ threshold.  
The $\psi(3770)$ decays primarily to $D^+D^-$ and $D^0\bar{D}^0$ 
pairs; it lacks sufficient energy to produce even one additional pion, 
which is the lightest hadron.  
It is common to reconstruct one $D$ meson in a well-understood 
hadronic final state (the ``tag'' side), and then study the decay 
of the other meson (the ``signal'' side) to some final state of interest.  
This tagging technique removes non-resonant collision events and also 
reduces combinatorics, i.e., the number of ways of forming the 
desired final state from the detected particles.  
MARKIII pioneered the use of $D$ tagging to measure absolute 
$D$ meson branching fractions \cite{MARKIIIhad, MARKIIIhadrev}; 
constrained kinematics also permit studies of final states with neutrinos.  



\subsection{Studies of the  $\psi(3770)$}
Properties of the $\psi(3770)$ resonance itself have long been of interest 
and BESII was an important contributor in this area.  
Using a sample of 27.7 pb$^{-1}$ taken near 3773 MeV, 
BESII provided the first evidence for a specific non-$D\bar{D}$ decay 
of this state: $\psi(3770) \to J/\psi \pi^+\pi^-$ \cite{3770Jpsipipi}.  
A signal of about 12 events  with a significance of more than 
three standard deviations indicated a branching ratio of order 0.3\%.  
By now, several more exclusive non-$D\bar{D}$ modes are known \cite{pdg}, 
but their sum is still only 0.5\%.  
One can investigate instead the inclusive, or total, 
non-$D\bar{D}$ branching fraction; 
two subsequent BESII papers addressed this issue.  
One measured the $D^0\bar{D}^0$, $D^+D^-$, and total hadronic cross sections 
vs.~CM energy across the $\psi(3770)$ peak region \cite{3770nonDD}.  
By subtracting the $D\bar{D}$ sum from the total, one obtains 
${\cal B}(\psi(3770) \to {\rm non-}D\bar{D}) = (16.4 \pm 7.3 \pm 4.2)\%$.  
An alternative analysis \cite{3770nonDD2} determined the 
total hadronic cross-section with 17.3 pb$^{-1}$ of data taken 
near the $\psi(3770)$ peak.  
Combined with previous determinations of $D^0\bar{D}^0$ and $D^+D^-$ 
peak cross sections \cite{3770DD}, BESII obtained 
${\cal B}(\psi(3770) \to {\rm non-}D\bar{D} = (14.5 \pm 1.7 \pm 5.8)\%$.  
There is a mild disagreement with a contemporaneous result from CLEO-c, 
which gave ${\cal B}(\psi(3770) \to {\rm non-}D\bar{D}) < 9\%$ 
at 90\% confidence level \cite{CLEOnonDD}, a limit extracted 
from a result with a central value quite close to zero.  
More precise determinations are desirable, but controlling systematic 
uncertainties is challenging.  

\subsection{Precision Semileptonic and Leptonic $D$ Decays}
BESII also performed a measurement of the semileptonic $D^0$ decays 
$D^0 \to K^- e^+ \nu_e$ and $D^0 \to \pi^- e^+ \nu_e$ \cite{semilep2004}.  
These measurements are of great interest as a middle-ground between 
all-hadronic final states (easy to measure, theoretically difficult) 
and all-leptonic decays (hard to measure, theoretically clean).  
Hadronic uncertainties are summarized as functions of $q^2 = m_{e\nu}^2$, 
known as form-factors (FF).  
Theory can provide a controlled series expansion of the form-factor shape 
\cite{BecherHill}, 
but the normalization has only been addressed by Lattice QCD (LQCD).  
One can use LQCD as an input, and directly extract 
the CKM matrix quark-couplings, $|V_{cd}|$ and $|V_{cs}|$.  
\begin{marginnote}[0.1in]
\entry{CKM MATRIX ${V_{qq'}}$}
{Measures the relative amplitude of 
 weak interaction transitions between quark types ${q}$ 
 and ${q'}$.
}
\end{marginnote}
This provides a valuable alternative to other CKM element determinations, 
which often assume there are no quarks beyond the six types currently known.  
One can also test LQCD FF shapes and, using external $|V_{cq}|$ values, 
also the FF normalizations (in particular, the ``intercepts,'' 
or values at $q^2 = 0$).  
The BESII work was the first threshold semileptonic measurement 
in fifteen years, since MARKII's initial work with 9.56 pb$^{-1}$ 
\cite{MARKIIIsemilep}.  
Modest statistics meant that only estimates of the form-factor intercepts,
$f_K^{+}(0)$ and $f_{\pi}^{+}(0)$, were obtained, 
by assuming naive FF shapes (from so-called ``pole models'').  
But this analysis was a bridge to the modern era: it was soon followed 
by higher-statistics CLEO-c results, which were then surpassed 
by BESIII.  

\begin{figure}[h]\centering
  \includegraphics[width=\linewidth]{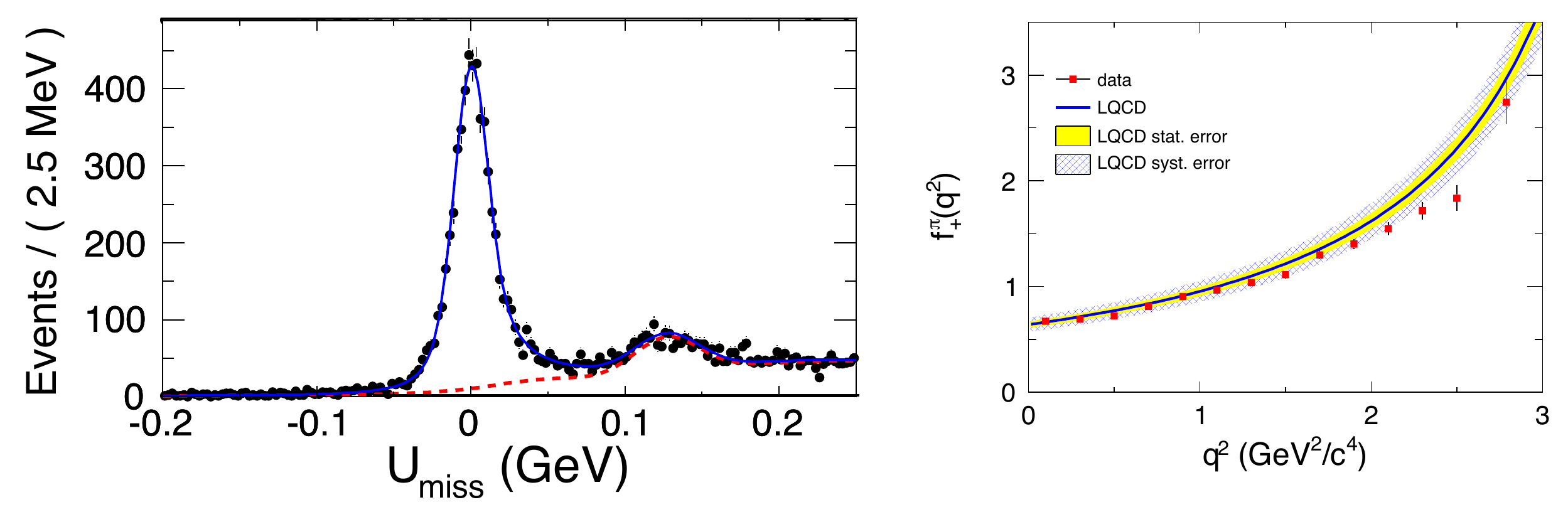}
  \caption{BESIII analysis of $D^0 \to \pi^-e^+\nu_e$.  
           Left: $U_{miss} = E_{miss} - p_{miss}$ distribution; 
           the blue curve is a fit to the data points, 
           including the red dashed background contribution             
           Right: The extracted form factor, $f_\pi(q^2)$, 
           compared to lattice QCD.  
           Modified from Ref. \cite{semilep2015} with permission. }
     \label{Fig:pienu}
\end{figure}

BESIII has accumulated 2.9 fb$^{-1}$ of data at the $\psi(3770)$ 
\cite{3770Lumi}, 
which is 3.5 times the previous largest sample, obtained by CLEO-c.  
With this, BESIII obtained the most precise semileptonic form factors 
obtained to date \cite{semilep2015}.  
In Figure \ref{Fig:pienu}, we display both the main signal plot 
for the more difficult (Cabibbo-suppressed) $\pi^-e^+\nu_e$ mode, 
\begin{marginnote}[0.1in]
\entry{Cabibbo Suppression}
{Reduction of decay rates due to CKM matrix elements much smaller than one.
}
\end{marginnote}
as well as the extracted form factor compared to a LQCD result.  
The key signal variable is $U_{miss} = E_{miss} - p_{miss}$, where 
``miss'' refers to the missing neutrino four-vector inferred 
via energy-momentum conservation.  
As with the following $D^+ \to \mu\nu$ analysis, it is impressive how 
clean a signal is obtained for suppressed decays involving undetectable 
neutrinos!  

\begin{figure}[h]
  \includegraphics[width=8.cm]{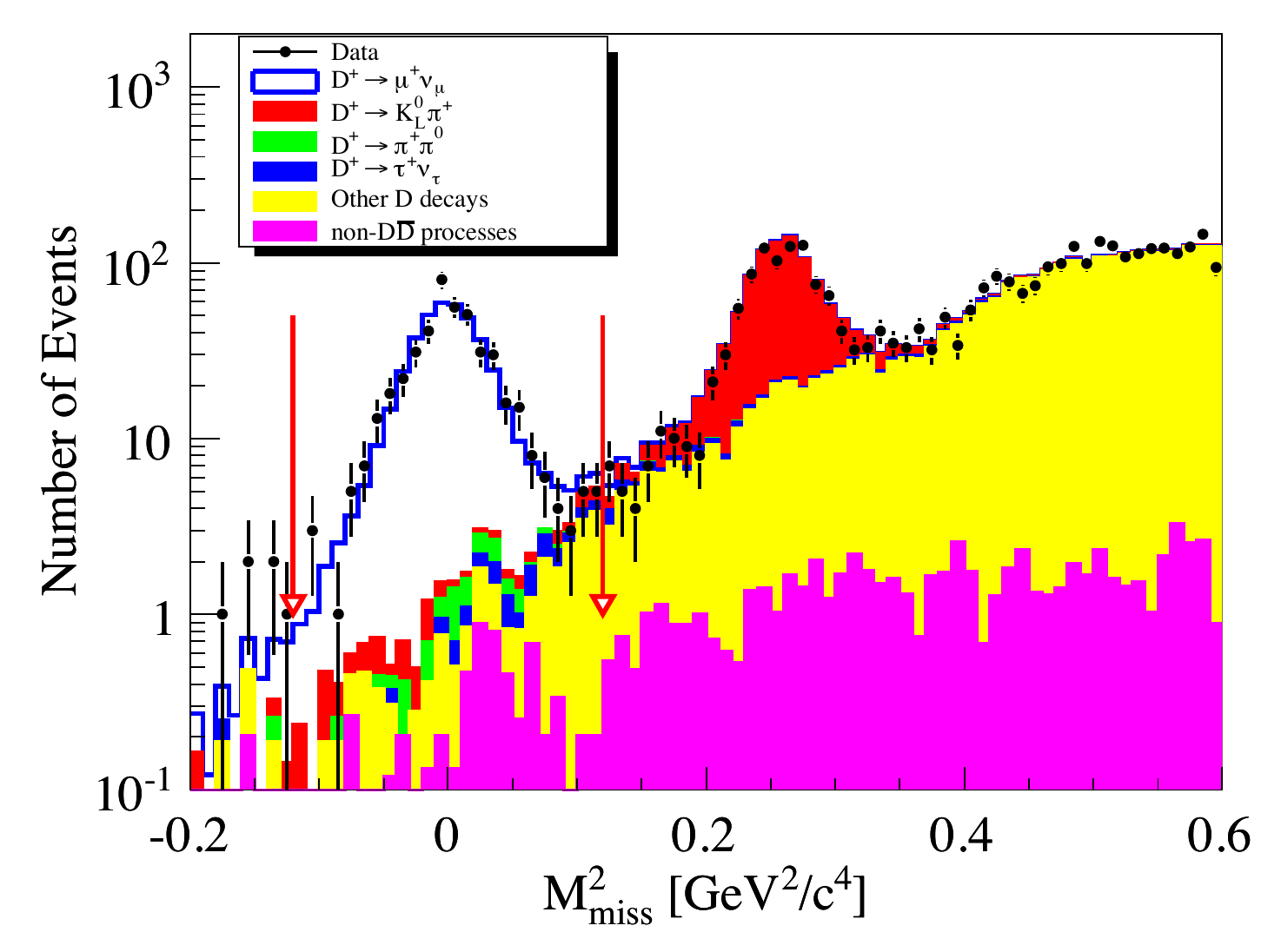}
  \caption{Distribution of missing-mass-squared, $M_{miss}^2$, 
           for the $D \to \mu \nu_\mu$ analysis, 
           showing clear and clean excess at $m_\nu^2 = 0$.  
           The blue fit to the data points includes many background 
           processes shown as colored shaded regions.  
           Modified from Ref. \cite{Dmunu} with permission. }
\label{Fig:munu}
\end{figure}

The purely leptonic decay $D^+ \to \mu^+ \nu_\mu$ is 
important since all hadronic uncertainties are summarized in one number, 
the pseudoscalar decay constant, $f_{D^+}$ \cite{PDG_DecaysOfCPS}.  
This is related to the square of the wave-function of the 
quarks forming the $D^+$ meson.  
%
It is experimentally challenging, due to Cabibbo-suppression of the rate 
and the presence of only one detectable decay product.  
The analysis reconstructs a hadronic $D$ tag opposite the signal decay, 
which here consists of only a single muon track.  
The final signal plot makes use of the missing-mass-squared, $MM^2$, 
calculated from the inferred neutrino four-vector, 
which should peak at $MM^2 = m_\nu^2 = 0$.  
The BESIII result is shown in Figure \ref{Fig:munu}.  
A well-known theoretical expression relates the branching ratio 
to the decay constant, and BESIII obtains 
$f_{D^+} = (203.2 \pm 5.3 \pm 1.8)$ MeV \cite{Dmunu}.  
This is the most precise determination to date, and it compares 
well to LQCD calculations \cite{PDG_DecaysOfCPS}.

\begin{textbox}[h]
\section{Quantum Correlations}
The pair of $D$ mesons produced in $\psi(3770)$ decays are correlated: 
the state of one influences the state of the other.  In particular, 
if one is detected decaying to an odd eigenstate of $CP$, then the other 
one must be even, and vice-versa.   
This involves the same basic quantum mechanics as the famous 
Einstein-Podolsky-Rosen correlations of photon pairs.  
\end{textbox}
\subsection{$CP$ Tagging of $D^0\bar{D}^0$ Pairs from the $\psi(3770)$}
Due to quantum correlations, charm threshold data allows for $CP$-tagging 
of neutral $D$ mesons: reconstructing one $D$ in a $CP$ eigenstate 
projects the other into the opposite $CP$ eigenstate.  
These eigenstates are linear superpositions 
of the form $(D^0 \pm \bar{D}^0)/\sqrt{2}$.  
Interference in  the decays of such states allows the extraction 
of relative phase information between $D^0$ and $\bar{D}^0$ decays.  
A measurement of the $D^0 \to K^-\pi^+$ ``strong phase'' difference 
has been performed in this manner \cite{Kpiphase} .  
This phase arises from strong-interaction scattering 
of the final-state particles.  
\begin{marginnote}[2.in]
\entry{$CP$}
{An operation wherein particles and antiparticles are interchanged ($C$), 
and left and right are inverted ($P$).
}
\end{marginnote}
BESIII directly measures the difference between the $CP-$ and $CP+$ 
eigenstates decaying to $K^-\pi^+$ divided by their sum, obtaining 
the asymmetry ${\cal A}_{CP} = (12.7 \pm 1.3 \pm 0.7)\%$.  
Combining this result with certain external inputs 
allows them to obtain 
$\cos\delta_{K\pi} = 1.02 \pm 0.11 \pm 0.06 \pm 0.01$ 
(errors are statistical, systematic, and external), indicating 
a small phase $\delta_{K\pi}$.  This result is useful in the 
interpretation of $D^0-\bar{D}^0$ oscillation results obtained 
using the $K\pi$ final state \cite{PDG_D0mix}.  
Other quantum-correlation analyses are underway; many of these 
are useful inputs to studies of the CKM matrix performed 
with $B$ meson decays.

\subsection{Beyond the $D^+$ and $D^0$}
BESIII continues to broaden its impact on charm physics with 
analyses using data from energies above the $\psi(3770)$.  
A recent measurement of ${\cal B}(\Lambda_c \to \Lambda e^+ \nu_e)$ 
\cite{Lambdac}, utilizes similar tagging techniques, but with 
$\Lambda_c\bar{\Lambda}_c$ pairs produced at 4.6 GeV. 
This technique has also been applied to hadronic final states, 
such as $\Lambda_c \to pK\pi$ \cite{Lambdac_had}, the ``golden mode'' 
which anchors most $\Lambda_c$ branching fractions.  
Another example is a recent precise measurement of the $D^{*0}$ 
branching fractions to $D^0 \pi$ and $D^0 \gamma$ \cite{DstarBFs}.  
Using 482 pb$^{-1}$ of data taken at $\sqrt{s} = 4009$ MeV, 
BESIII obtains the ratio of decay widths 
$\Gamma(D^{*0} \to D^0 \pi^0)/\Gamma((D^{*0} \to D^0 \gamma) 
= 1.90 \pm 0.07 \pm 0.05$, which is both more precise 
and noticeably higher than previous results.  

In the near future, BESIII anticipates dedicated running at 4170 MeV, 
where $D_s^{*\pm}D_s^\mp$ pairs are produced in abundance, 
thus adding precision $D_s$ physics to the BESIII charm portfolio.  
In addition, significant samples exist at a variety of other open-charm 
energies, as described above in the discussion of the $XYZ$ states.  
With this wealth of data, one may expect not only increased precision 
on existing results from charm threshold, but also novel uses 
of the large and varied datasets of BESIII.

\section{FUTURE PROSPECTS}

The scientific output summarized in this review is both broad in scope 
and accelerating in pace.  
Physics results obtained by BES inform a variety of subjects, directly 
impacting our understanding of both weak and strong interactions.  
Traditional areas of strength have been supplemented by new topical areas, 
such as studies of the exotic $XYZ$ states and searches for new particles.  

The current BESIII collaboration has grown, 
both in size and in international participation, 
into a leading player in particle physics today.  
Existing data sets are proving to be very productive and data-taking runs 
are anticipated to continue beyond 2020.  
As discussed in the previous sections, these future physics results 
promise to be an substantial addition to the existing BES legacy.  
The possibility of unexpected surprises only reinforces our appreciation 
for those who worked to first build this program many years ago.


\newpage

\section*{DISCLOSURE STATEMENT}
The authors are not aware of any affiliations, memberships,
funding, or financial holdings that might be perceived as affecting
the objectivity of this review.

\section*{ACKNOWLEDGMENTS}
The authors wish thank their BESI/BESII/BESIII collaborators for their
great efforts. One author (FAH) wishes to thank Prof. Zhipeng Zheng
for providing many details on the early history of BESI. The authors are
supported by the United States Department of Energy under contract
numbers DE-SC-0010504, DE-FG02-05ER41374, and DESC0010118.

\input{citations.tex}

\end{document}

%% file: citations.tex
\bibliographystyle{<.bst ar-style5.bst>}